\documentclass{aastex63}

\submitjournal{ApJ}

\shorttitle{}
\shortauthors{Zeng \& Yang}

\graphicspath{{./}{figures/}}

\begin{document}

\title{Oceanic Superrotation on Tidally Locked Planets}

\correspondingauthor{Jun Yang}
\email{junyang@pku.edu.cn}

\author[0000-0002-2624-8579]{Yaoxuan Zeng}
\affiliation{Department of Atmospheric and Oceanic Sciences, School of Physics, Peking University, Beijing 100871, China.}

\author[0000-0001-6031-2485]{Jun Yang}
\affiliation{Department of Atmospheric and Oceanic Sciences, School of Physics, Peking University, Beijing 100871, China.}


\begin{abstract}

Is there oceanic superrotation on exoplanets? Atmospheric superrotation, characterized by west-to-east winds over the equator, is a common phenomenon in the atmospheres of Venus, Titan, Saturn, Jupiter, and tidally locked exoplanets. The stratospheric atmosphere of Earth is also superrotating during the westerly phase of the quasi-biennial oscillation (QBO). However, whether the same phenomenon can occur in ocean is poorly known. Through numerical simulations, here we show that oceanic superrotation does occur on tidally locked terrestrial planets around low-mass stars. Its formation (spun-up from rest) is associated with surface winds, the equatorward momentum convergence by Rossby waves, and the eastward propagation of Kelvin waves in the ocean. Its maintenance is driven by equatorward momentum transports of coupled Rossby-Kelvin waves in the ocean excited from the uneven stellar radiation distribution. The width of the superrotation is mainly constrained by the Rossby deformation radius in the ocean, while its  strength is more complex. Many factors can influence the strength, including planetary rotation rate, stellar flux, greenhouse gas concentration, seawater salinity, bottom drag, and a scaling theory is lack. This work confirms that superrotation can occur on tidally locked terrestrial planets with seawater oceans and suggests that it may also occur on tidally locked hot planets with magma oceans that will possibly be observed in the near future.
\end{abstract}


\section{Introduction} \label{sec:intro}

Atmospheric superrotation is characterized by eastward wind at the equator, which means the atmosphere there has a higher angular momentum than the solid surface. Atmospheric superrotation is a common phenomenon across the universe. In the solar system, superrotation exists in the atmospheres of Venus, Titan, Saturn, Jupiter as well as the stratospheric atmosphere of Earth during the westerly phase of the quasi-biennial oscillation (QBO) (e.g. \cite{kraucunas2005equatorial, schneider2009formation, read2018superrotation, lutsko2018response}). In order to maintain atmospheric superrotation, there must be momentum transports from higher latitudes to the equator against friction or other processes, according to angular momentum conservation (\cite{hide1969dynamics,held1999equatorial,showman2013atmospheric}). This up-gradient transport into the jet can result from Rossby waves, coupled Rossby-Kelvin waves, mixed Rossby-gravity waves, wave-jet resonance, barotropic instability, or baroclinic instability (\cite{suarez1992terrestrial,del1996simulations,joshi1997simulations,lee1999climatological,williams2003barotropic, kraucunas2005equatorial, schneider2009formation,caballero2010spontaneous,mitchell2010transition,showman2007atmospheric,showman2010matsuno, showman2011equatorial,liu2011convective,arnold2012abrupt,tsai2014three, pinto2014atmospheric,wang2014planetary, laraia2015superrotation,read2018superrotation,lutsko2018response,pierrehumbert2019atmospheric}). For example, \cite{kraucunas2005equatorial} suggested that in an Earth-like atmosphere, equatorial superrotation can be generated by equatorward stationary eddy momentum convergence, which is associated with zonal variations in the diabatic heating at low-latitudes. \cite{mitchell2010transition} studied the transition from current Earth-like atmospheric circulation to an equatorial superrotation state. They found that during the spin-up period, superrotation is generated by equatorward momentum convergence associated with both barotropic and baroclinic instabilities.

On tidally locked planets like hot Jupiters, there also exists superrotation in the atmosphere. \cite{showman2011equatorial} applied a two-layer shallow water model to study the maintenance mechanism of the superrotation on tidally locked planets. They found that the atmospheric superrotation on tidally locked planets is maintained by the momentum convergence by stationary eddies. Such stationary eddies are associated with the equatorial Rossby-Kelvin waves in the Matsuno-Gill mode (\cite{matsuno1966quasi, gill1980some}). These waves tilt northwest-southeast in the northern hemisphere (and northeast-southwest in the southern hemisphere) and are ultimately generated by the day-night stellar radiation contrast. \cite{tsai2014three} further applied a 3D model to study the vertical structure of the atmospheric superrotation. They found that the tilting vertical eddies can result in a vertical eddy momentum transport, which contributes to the vertical structure of the equatorial eastward jet on hot Jupiters.

Atmospheric superrotation has a profound effect on the observation of the thermal phase curve of tidally locked planets. Stellar radiation is strongest at the substellar point on tidally locked planets. However, the hottest spot is shifted eastward of the substellar point due to the Doppler shifting of the phase of the stationary wave caused by the westerly equatorial jet (\cite{showman2010matsuno, showman2011equatorial, showman2013atmospheric, tsai2014three, hammond2018wave}). As a result, the peak of the phase curve is shifted eastward, which was first suggested by \cite{showman2002atmospheric}, and was then confirmed by observations on hot Jupiters (\cite{knutson2007map, zellem20144}).

Although the maintenance mechanism of atmospheric superrotation, as well as its effect on observation and climate is relatively well-known, whether and how superrotation can occur in another geophysical fluid--ocean has been poorly studied. Oceanic superrotation does exist under several circumstances on Earth. In the Pacific and Atlantic Oceans, there are year-round equatorial currents flowing from west to east, which are known as ``Equatorial Undercurrent'' (\cite{cromwell1954equatorial,philander1973equatorial}). They are driven by west-to-east pressure gradient forces associated with the surface trade winds and by inertial dynamics related to extra-equatorial inflows under potential vorticity conservation (\cite{charney1959non,charney1971structure,philander1980generation,philander1980equatorial,vallis2006atmospheric}). Although these currents can be called as oceanic superrotation, their spatial scales, about 200 m in the vertical direction and about 400 km in the meridional (south-north) direction, are too small to get much attention in studying planetary climate.

On tidally locked planets, there may also exist superrotation in the ocean. Recent numerical simulations with fully coupled atmosphere-ocean general circulation models showed that there are also west-to-east currents in the tropics of tidally locked terrestrial planets (\cite{hu2014role,yang2014water,yang2019ocean,yang2020transition,del2019habitable}). These currents can transport heat from the substellar region to the night side, acting to warm the night side and subsequently aid to avoid atmospheric collapse (i.e., CO$_2$ condenses down to the surface) or water trapping (i.e., all the water freezes on the night side surface) (\cite{hu2014role,yang2014water}). However, none of the previous studies had realized that the currents belong to equatorial superrotation, and the formation and maintenance mechanisms of the currents were unknown.

The goal of our study is to show the phenomenon of oceanic superrotation on tidally locked planets, and to uncover the formation and maintenance mechanisms of the equatorial westerly jet. We also aim to examine the dependence of oceanic superrotation on rotation rate, as well as to study its effect on climate and observation. The structure of this paper is as follows. Section \ref{sec:Methods} introduces the models we apply in this work and experiment designs. Section \ref{sec:superrotation} demonstrates the oceanic superrotation phenomenon on tidally locked planets. Section \ref{sec:formation} and \ref{sec:maintenance} address the formation and the maintenance mechanisms, respectively. Section \ref{sec:rotation} shows the dependence of the superrotation on planetary rotation rate and how it influences the planetary climate. Section \ref{sec:summary} is the summary and discussions.

\section{Methods}\label{sec:Methods}

We use the three-dimensional (3D) global climate model--the Community Climate System Model version 3 (CCSM3, \cite{collins2006community, rosenbloom2011using}) to simulate the oceanic circulation on tidally locked terrestrial planets. The model has four coupled components: atmosphere, land, ocean, and sea ice. The ocean part is the Parallel Ocean Program version 1.4 (POP1), developed by the Los Alamos National Laboratory (\cite{smith2002reference}). A relatively low resolution of CCSM3 and the Gent-McWilliams parameterization for mesoscale eddies in the ocean (\cite{gent1990isopycnal}) are used in our control simulation. The atmosphere and land components of the model have a horizontal resolution of 3.75$^\circ$\,$\times$\,3.75$^\circ$, with 26 vertical levels from the surface to around 36 km. The ocean and sea-ice components have a variable latitudinal resolution starting at 0.9$^\circ$ near the equator to 3.6$^\circ$ near the poles, a constant longitudinal resolution of 3.6$^\circ$, and 25 vertical levels. Due to the slow rotation of the planets, the Rossby deformation radius is larger than the model resolution, thereby mesoscale eddies in the ocean can be roughly resolved in the simulations (\cite{hu2014role, del2019habitable, yang2020transition}). Note that in the model there are no grid points right on the equator, so equatorial values shown in the figures of this paper are obtained by averaging the values on both sides of the equator.


\begin{figure}[b]
\centering
\includegraphics[width=0.86\linewidth]{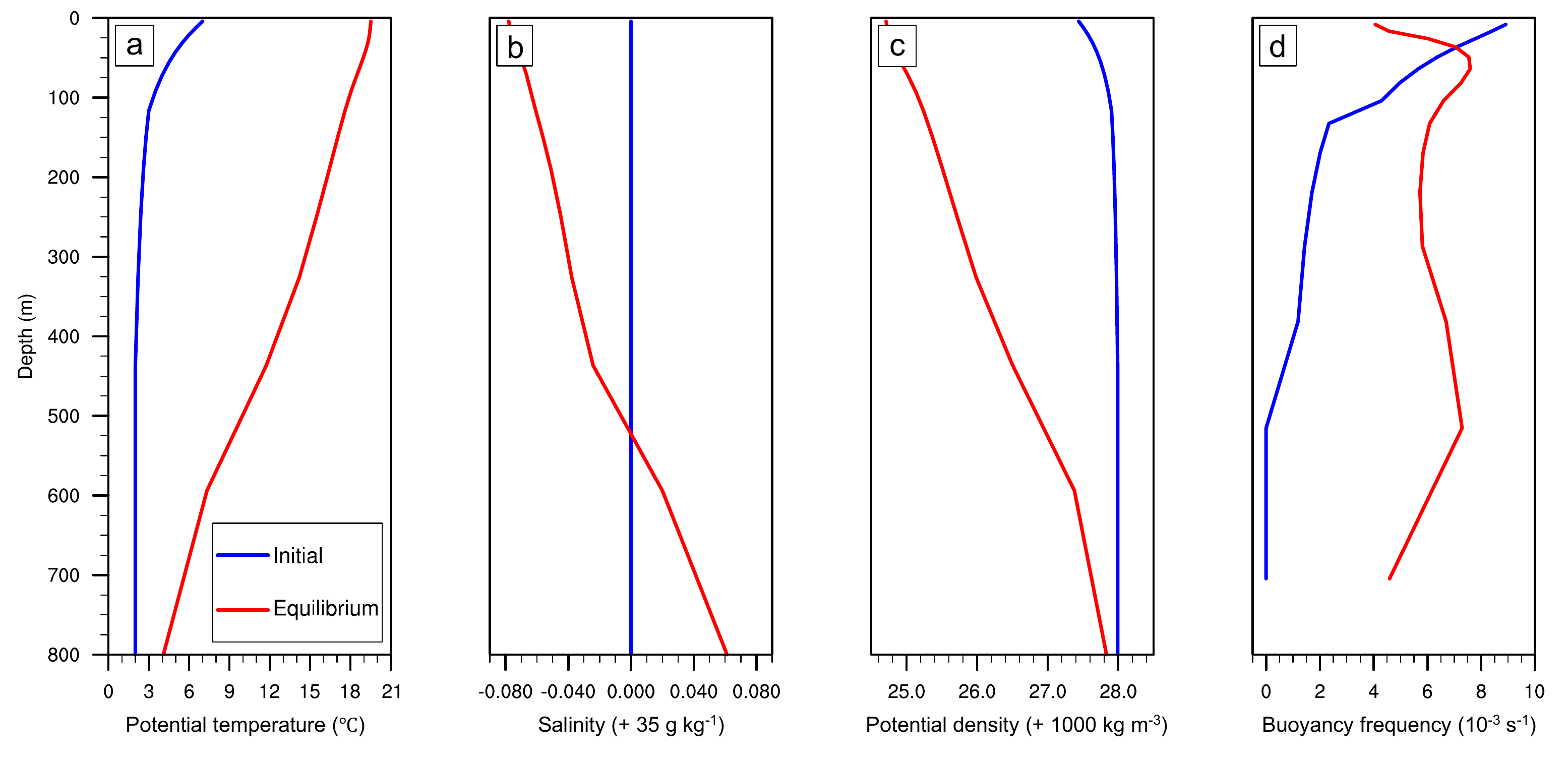}
\caption{Initial oceanic condition (blue lines) and equilibrated oceanic condition (red lines) in the control experiment. (a) potential temperature, (b) salinity, (c) potential density, and (d) Brunt-V\"ais\"al\"a frequency $N$. For the initial condition, all model grids have the same values at each layer. For the equilibrium condition, the red lines show the global-mean values.}
\label{figS_initial_equilibrium}
\end{figure}

By default, we employ an aqua-planet with an ocean depth of around 1000~m; sea-ice salinity is 4 g kg$^{-1}$ while seawater salinity is 35 g kg$^{-1}$; the atmosphere is set to be Earth-like; the rotation period ($=$ orbital period) is 30 Earth days; both obliquity and eccentricity are zero; other planetary parameters, such as radius and gravity, are equal to Earth’s values. In this paper, the length of one day is defined following Earth’s value, which is 24 hours. The stellar flux is set to 1400 W m$^{-2}$. Geothermal heat flux is set to zero, so the energy source is stellar radiation only. The stellar temperature is 4500~K; sea ice and snow albedos are lower than those on Earth due to the red shift of the stellar spectrum compared to the Sun. We have also carried out series of simulations to test the effects of rotation rate, gravity, salinity, and eddy parameterization. Five additional rotation periods have been examined, 8, 10, 60, 80, and 100 days. In all the experiments, the rotation period is equal to the orbital period, i.e., in synchronous rotation. For the brine rejection process during the formation and melting of sea ice, the ocean reference salinity and the sea ice reference salinity are used to calculate the salinity exchange between sea ice and seawater. For more detailed descriptions of the model and experimental designs, please read \cite{yang2019ocean, yang2020transition}.

When modifying the rotation period, we fixed the stellar flux as constant, similar to the methods applied in \cite{merlis2010atmospheric}, \cite{way2016venus}, \cite{Noda2017the}, and \cite{bin2018new}. This is different from \cite{Kopparapu2017the} and \cite{kopparapu2017habitable}, who modified the rotation period and the stellar flux simultaneously. By fixing the stellar flux while changing other parameters like rotation periods, we can clearly separate the effect of each parameter from others, although we cannot self-consistently consider the combined effects of these factors.


\begin{figure}[t]
\centering
\includegraphics[width=0.9\linewidth]{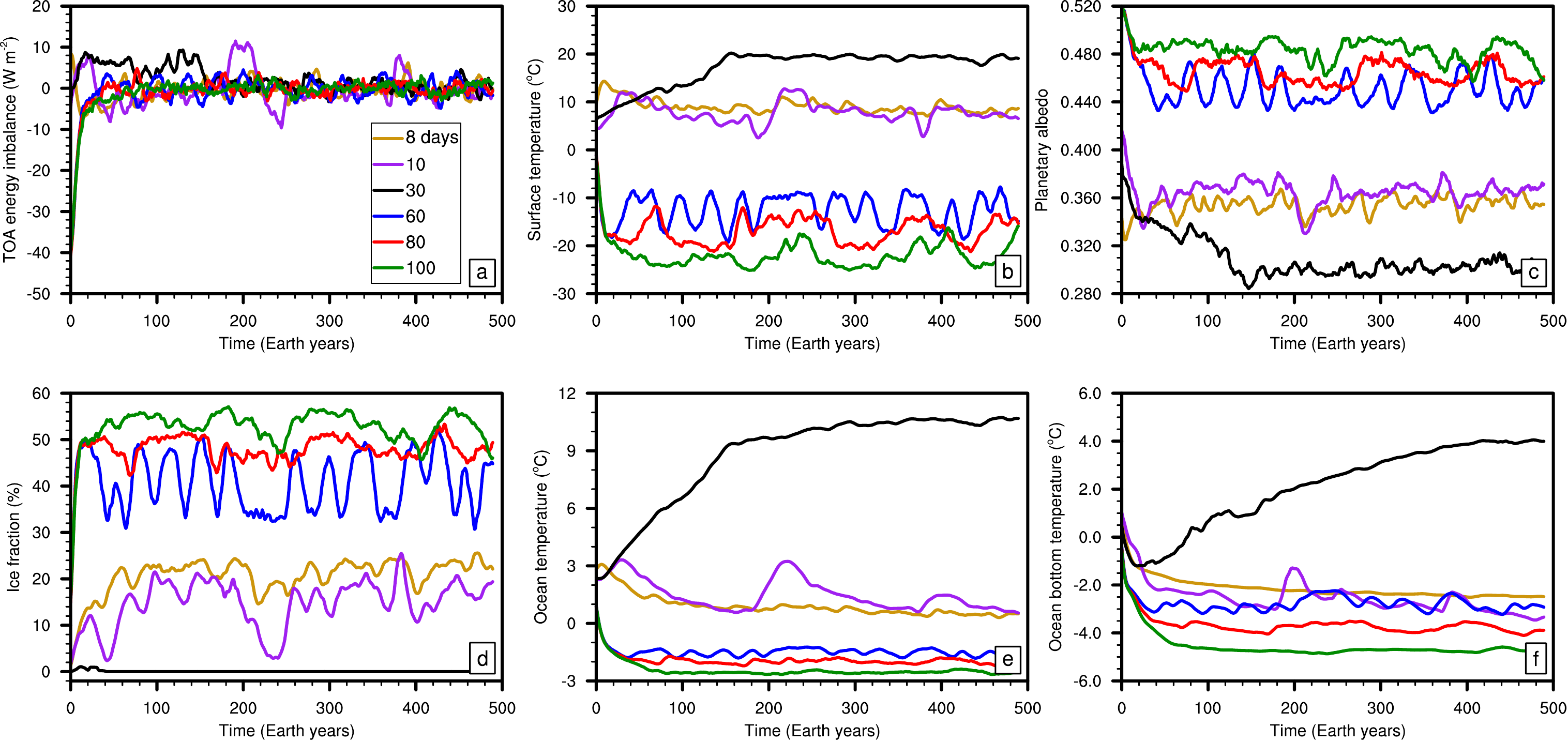}
\caption{Ten-year running average time series of global-mean energy imbalance at the top of the atmosphere (TOA, (a)), surface air temperature (b), planetary albedo (c), sea ice fraction (d), global ocean temperature (e), and ocean bottom temperature (f) in the coupled atmosphere--ocean experiments of varying rotation periods. The strong variability in the experiments is from the coupling among atmosphere, ocean, and sea ice, but the detailed mechanism(s) is unknown yet. In atmosphere-only experiments, there is no strong variability.}
\label{figS_timeseries}
\end{figure}

The atmosphere was initialised from a state close to modern Earth, and the ocean was initialised from a state of rest with a globally-uniform temperature and salinity but with a moderate stratification shown in Figure~\ref{figS_initial_equilibrium}. Each case was run for about 500 Earth years (Figure~\ref{figS_timeseries}), and averages of the last 100 years were used to analyze the mean climate and averages of monthly output were used to calculate momentum budgets.

To isolate the separate roles of surface winds and heat fluxes in driving the oceanic circulation, we also did experiments using the ocean component of the model--POP1. By default, the surface winds and heat fluxes are specified from the fully coupled atmosphere-ocean experiments; under which, the ocean-only model is able to well reproduce the oceanic circulation obtained in the coupled model. Through modifying the surface winds, we can know how the surface winds influence the oceanic circulation, and the similar way is applied to heat fluxes. Moreover, through changing the rotation rate but leaving the winds and heat fluxes unchanged, we can also know how the Coriolis force in the ocean influences the oceanic circulation.

To isolate the role of oceanic circulation in planetary climate, we preformed corresponding atmosphere-only experiments using the Community Atmosphere Model version 3.1 (CAM3, \cite{collins2004description}), which is the atmosphere component of CCSM3. CAM3 is coupled with a 50-m immobile ocean, and ocean heat transport is specified to be zero everywhere (\cite{yang2013stabilizing}). The model is coupled to a thermodynamic sea ice model, in which sea ice flows are not considered. Each case was integrated for 35 Earth years and reached a steady state after about 25 years. Averages over the last 5~years were used for our analyses.

\section{Oceanic superrotation} \label{sec:superrotation}

\begin{figure}[t]
\centering
\includegraphics[width=1.0\linewidth]{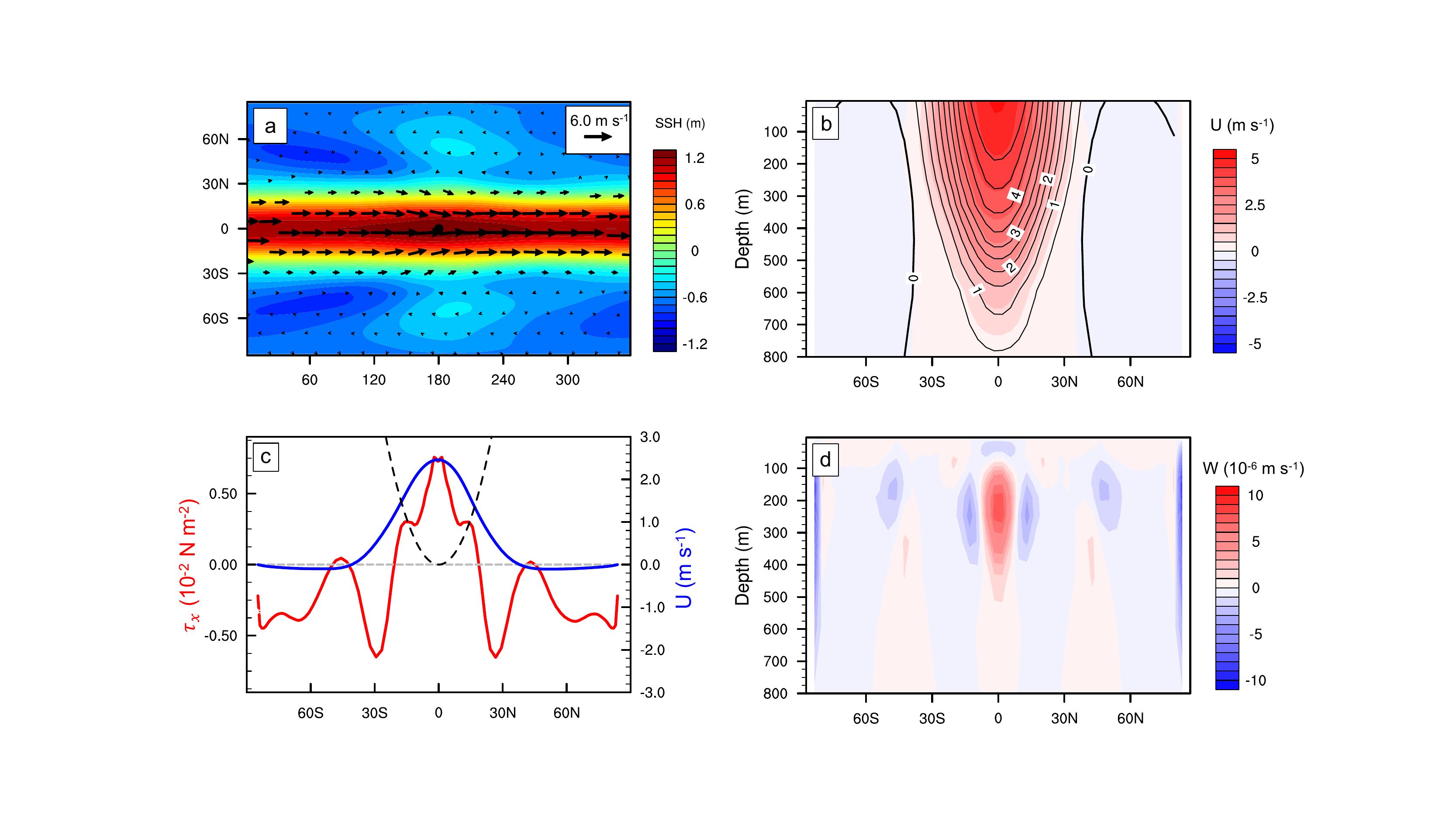}
\caption{Oceanic superrotation on a tidally locked aqua-planet. (a): surface ocean velocity (vector) and sea surface height (shading) as a function of longitude and latitude; (b): zonal-mean zonal ocean velocity as a function of latitude and depth; (c): zonal-mean zonal wind stress (red), vertically-averaged zonal ocean velocity (blue), and the zonal velocity following angular momentum conservation for a resting parcel that flows starting from the equator (black dashed); (d): same as (b) but for vertical velocity. In (a), the substellar point is marked with a black dot. In (b), the contour lines show the geostrophic velocity based on the balance between pressure gradient force and the Coriolis force, and the interval is 0.5~m\,s$^{-1}$. In this control experiment, the planetary rotation period ($=$ orbit period) is 30 Earth days, stellar flux is 1400~W m$^{-2}$, and CO$_2$ concentration is 300~ppmv.}
\label{fig2_current}
\end{figure}

In the control experiment with a rotation period ($=$ orbital period) of 30 Earth days, the main feature of the oceanic circulation is a strong current that flows from west to east in the region between around 40$^\circ$S and 40$^\circ$N (Figure~\ref{fig2_current}(a)~\&~(b)). This current is strong, wide, and deep. The speed of the current is several meters per second, its width is comparable to planetary size (around 80$^\circ$ in latitude), it encircles the entire equator, and in the vertical direction it extends from the surface to the seafloor. This is because the ocean bottom is set to be flat everywhere and subsequently form drag associated with ocean ridges (\cite{vallis2006atmospheric}) is absence. In the region between around 25$^\circ$S and 25$^\circ$N, this current has angular momentum higher than the solid surface at the equator (i.e., $M = (\Omega a cos\phi+ u(\phi))a cos\phi > \Omega a^2$, where $\Omega$ is planetary rotation rate, $a$ is planetary radius, $\phi$ is latitude, and $u$ is zonal ocean velocity), or equivalently has a velocity greater than that constrained by angular momentum conservation for a resting parcel that flows starting from the equator ($u(\phi)>\Omega a sin^2\phi/cos\phi$, Figure~\ref{fig2_current}(c)), so it can be called as ``oceanic superrotation''. 

\begin{figure}[t]
\centering
\includegraphics[width=0.8\linewidth]{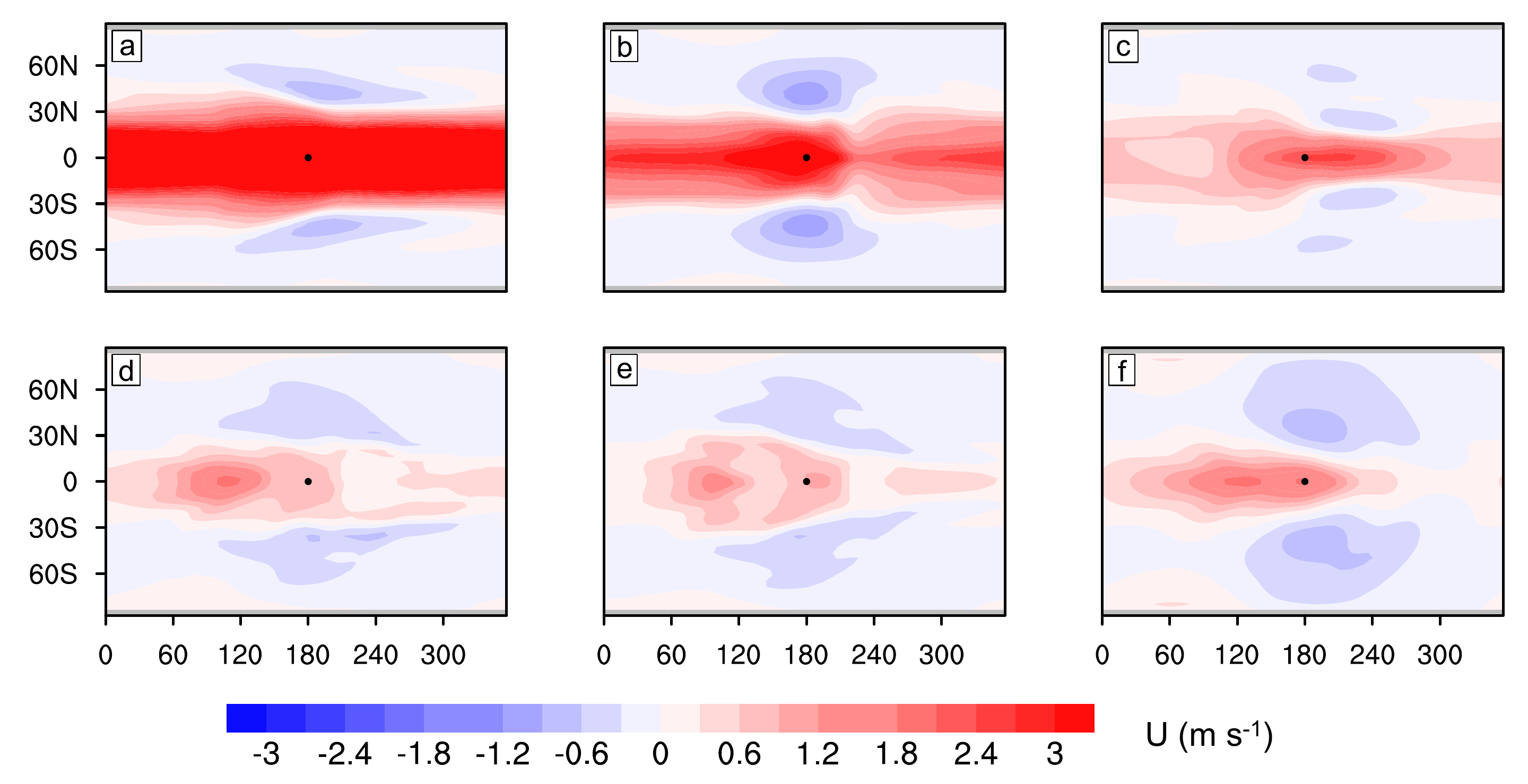}
\caption{Long-term mean zonal oceanic surface velocity in the sensitivity experiments. (a): the control experiment for comparison; (b): doubling the planetary gravity; (c): lowering the salinity (the seawater salinity is changed from 35 to 4 g kg$^{-1}$ and the sea ice salinity is changed from 4 to 0 g kg$^{-1}$); (d): Same as (c) but with a stronger brine rejection process in which the salinity difference between seawater and sea ice is artificially (unrealistically) set to 31 g kg$^{-1}$; (e): Same as (d) but the bottom drag coefficient is increased to 10 times of the default value; (f): Same as (d), but the parameterization of horizontal tracer diffusion is set to be the biharmonic horizontal diffusivity scheme instead of the default Gent-McWilliams parameterization (\cite{smith2002reference}). The black dots indicate the substellar point. In all these experiments, the planetary rotation period ($=$ orbit period) is 30 Earth days, stellar flux is 1400 W\,m$^{-2}$, and CO$_2$ concentration is 300 ppmv. Superrotation exists in all the experiments although its magnitude varies significantly.}
\label{figS_sensitivity}
\end{figure}

Oceanic superrotation is a robust phenomenon on tidally locked planets. When we modify planetary gravity, seawater salinity, sea ice salinity, and the parameterization scheme of oceanic mesoscale eddies, superrotation exists in all the experiments although its magnitude varies (Figure~\ref{figS_sensitivity}). Moreover, its existence does not depend on planetary size, atmospheric mass, greenhouse gas concentration, the strength of stellar flux, stellar spectrum, ocean depth, sea ice/snow albedo (see Figure~1 in \cite{hu2014role}, Figure~1 in \cite{yang2014water}, Figure~5 in \cite{yang2019ocean}, and Figure~S5 in \cite{yang2020transition}), or the global model employed (Figure~3 in \cite{del2019habitable}). Even in the experiments with land, there is superrotation in local regions as long as the day side is not mostly covered by continents (Figure~12 in \cite{yang2019ocean}). 


\begin{figure}[t]
\centering
\includegraphics[width=0.8\linewidth]{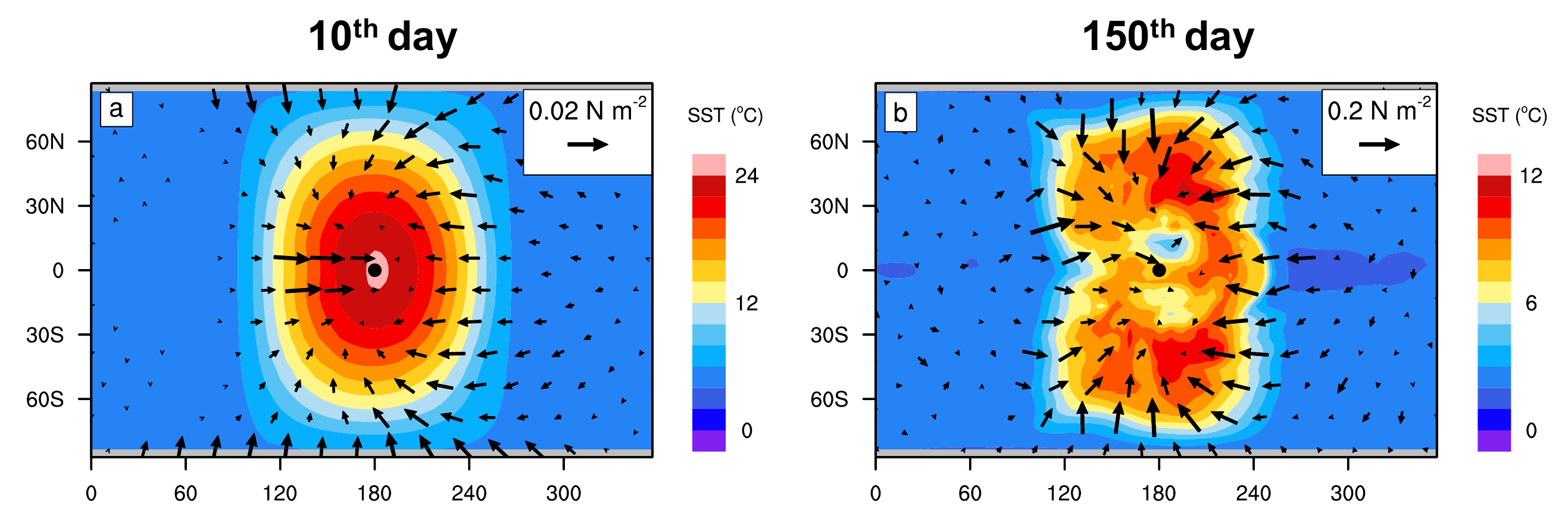}
\caption{Sea surface temperature (SST, color shading) and wind stress (vectors) on the 10$^{th}$ day (a) and the 150$^{th}$ day (b) of the control experiment. The black dots indicate the substellar point. Wind stresses smaller than $5\times 10^{-4}$ N m$^{-2}$ in (a) and $0.01$ N m$^{-2}$ in (b) are not shown. Note that the range of the color bar and the reference vector are different in these two panels. The wind stress on the 10$^{th}$ day is very small, especially on the night side due to almost uniform sea surface temperature there. However, there is still wind stress near the terminators where temperature gradient is large, and this wind stress can result in convergence of the ocean current towards the substellar point.}
\label{fign_initialwind}
\end{figure}

\section{Formation mechanism} \label{sec:formation}


\begin{figure}[t]
\centering
\includegraphics[width=1.02\linewidth]{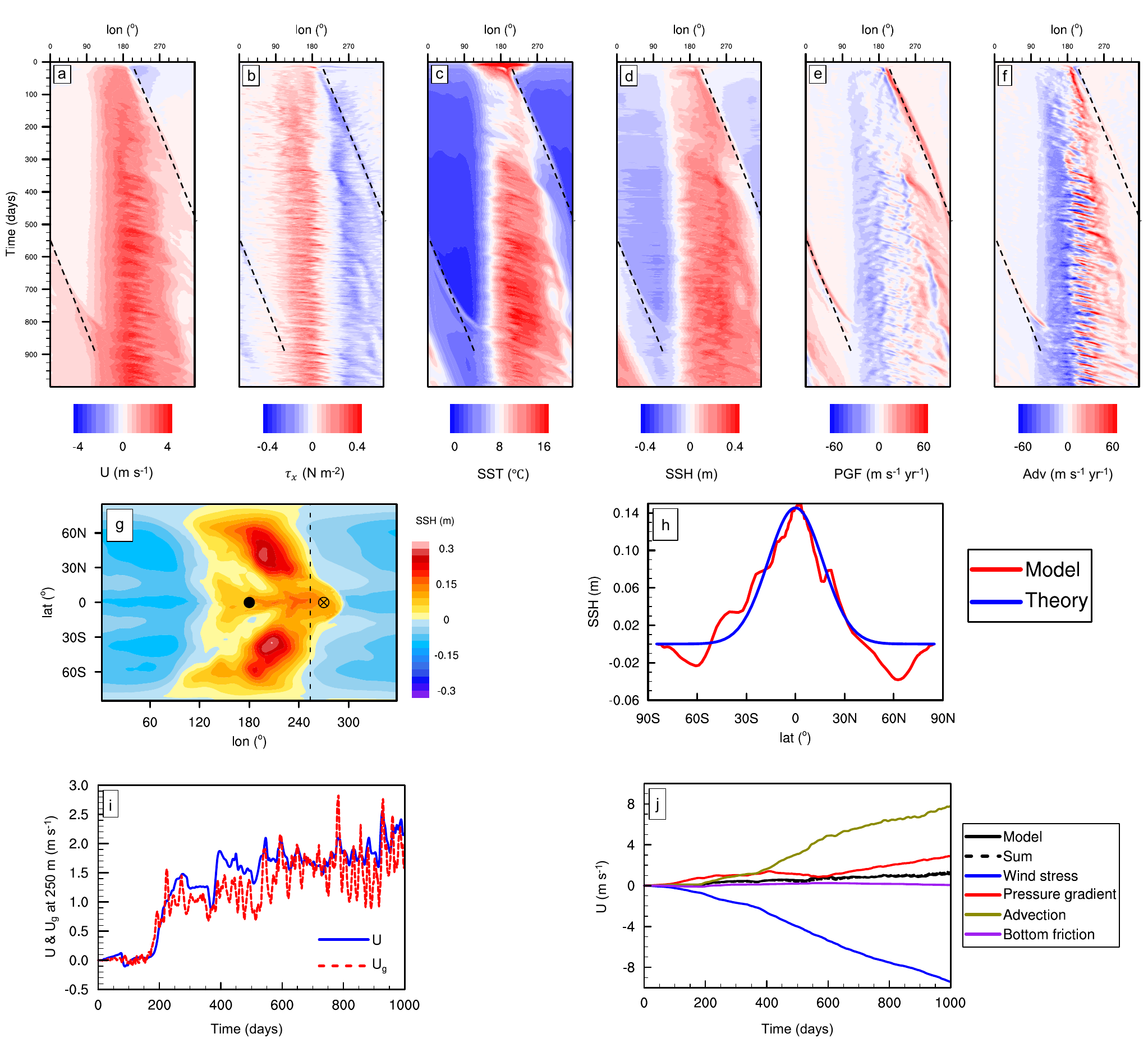}
\caption{Generation of the oceanic superrotation during the spin-up period in the control experiment. Evolution of the surface zonal ocean velocity (a), zonal wind stress (b), sea surface temperature (SST, c), sea surface height (SSH, d), accelerations due to zonal pressure gradient force (PGF, e), and accelerations due to nonlinear advection (f) along the equator between 2$^{\circ}$S and 2$^{\circ}$N. The dashed lines in (a)-(f) are the wave front of the initially excited Kelvin wave which is the main process for the adjustment of the ocean. (g): Sea surface height anomaly (zonal-mean values have been subtracted) on the 260$^{th}$ day; (h): Comparison between the model output and the analytic solution of the Kelvin wave for sea surface height anomaly at the longitude of 254$^{\circ}$ (vertical dashed line in panel (g)); (i): comparison for the zonal velocity at one point at the equator ($\otimes$ in the panel (g)) between the model output (blue) and that constrained by the meridional geostrophic balance (red). (j): Contributions to the vertically averaged zonal velocity (black solid) from wind stress (blue), pressure gradient force (red), nonlinear advection (dark yellow), bottom friction (purple), and the sum of the terms above (black dashed) at the same point as in (i). The black dot in (g) indicates the substellar point. Note that the black dashed line in (j) is almost overlaid by the black solid line. The experiment was initialized from a rest state with vertical stratification shown in  Figure~\ref{figS_initial_equilibrium}.}
\label{fig3_generation}
\end{figure}

In this section, we address the formation process, i.e., the response of a stratified ocean, initially at rest, to a sudden onset of the stellar radiation and planetary rotation. The wind stress develops first due to thermal contrast (Figure \ref{fign_initialwind} \& Figure \ref{fig3_generation}(b)) and then drives ocean currents because the inertia of air is smaller than that of seawater, so that the response of the atmosphere to the stellar radiation is much faster than that of the ocean. The surface wind stresses are very weak during the first several days especially on the night side, because surface temperature gradients are small there (Figure~\ref{fign_initialwind}(a)). After the first tens of days, the wind stress pattern--converging towards the substellar point generally develops (Figure~\ref{fign_initialwind}(b) \& Figure~\ref{fig3_generation}(b)). The rough pattern of the surface winds does not change much after it has been established, although it moderately evolves with time as temperature gradient evolves. 

As a result of the converging wind stress field, the oceanic zonal velocity is eastward on the west of the substellar point but westward on the east of the substellar point, on the day side (Figure~\ref{fig3_generation}(a)). The eastward current on the west of the substellar point gradually expands to the east following the propagation of Kelvin waves. Due to the convergence of surface winds and the thermal forcing from the stellar radiation (Figure~\ref{fig3_generation}(c)), sea surface height in the substellar region is higher than the surrounding (Figure~\ref{fig3_generation}(d) \& (g)). This disturbs the ocean and excites Rossby waves and equatorial Kelvin waves (\cite{philander1984unstable}). The Kelvin waves propagate eastward and transport momentum to the east through establishing positive zonal pressure gradient forces (Figure~\ref{fig3_generation}(d)~\&~(e)) as well as nonlinear advection accelerations (Figure~\ref{fig3_generation}(f)~\&~(j)). The phase speed ($c_p$) of the equatorial Kelvin wave is equal to $NH$, where $H$ is the ocean depth within which the wave propagates (around 400 m in the control experiment) and $N$ is the Brunt-V\"ais\"al\"a frequency,  which is a measure of the fluid stratification (10$^{-3}$ s$^{-1}$ in the experiment, see Figure \ref{figS_initial_equilibrium}(d) at the level of 400 m). As a result, the value of $c_p$ is about 0.4 m s$^{-1}$. This wave propagates from the substellar point to the east and then to the entire night side. This process takes about 900 Earth days, within which the zonal current along the entire equator becomes eastward (Figure \ref{fig3_generation}(a)).

During the formation period, the meridional (south-north) shape of the sea surface height at the wave front well follows the analytical solution of equatorial Kelvin waves in the form of $exp{(-\beta y^2 / 2c_p)}$, where $\beta$ is the northward gradient of the Coriolis parameter and $y$ is the northward distance from the equator (Figure \ref{fig3_generation}(h)). The speed of the zonal current can be estimated roughly using the meridional geostrophic balance near the equator, which is $u_g=-(g/\beta y)(\partial \eta / \partial y)$, where $g$ is the gravity and $\eta$ is the sea surface height (red line in Figure \ref{fig3_generation}(i), as well as the black lines in Figure~\ref{fig2_current}(b) for equilibrium state).


\begin{figure}[t]
\centering
\includegraphics[width=1.0\linewidth]{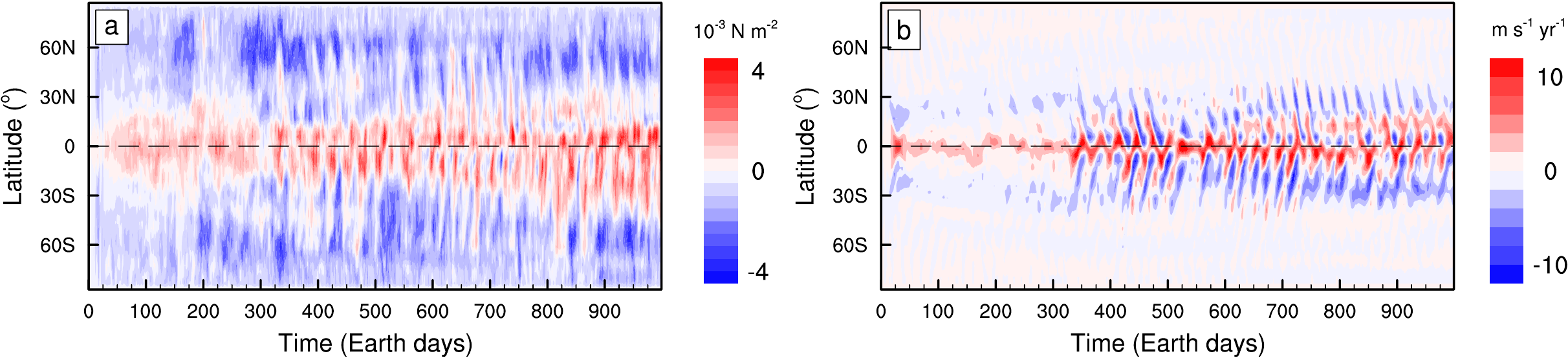}
\caption{Time series of the first 1000 days of the control experiment for the zonal-mean zonal wind stress (a) and the zonal-mean zonal acceleration due to horizontal eddies ($-\partial ([u^* v^*]) / \partial y$ where the bracket and asterisk indicate zonal average and deviation, respectively) in the first layer of the ocean. The black dashed lines indicate the equator.}
\label{fign_formation_rossby}
\end{figure}

The Kelvin wave alone cannot generate superrotation--it can only transport momentum along the equator to the night side, so that there must be other mechanisms for the equatorward momentum convergence. Besides the effect of the eastward zonal-mean zonal wind stress (Figure \ref{fign_formation_rossby}(a)), Rossby waves are evident at both low-latitudes and high-latitudes (Figure \ref{fig3_generation}(g)). The Rossby waves near the equator converge momentum to the equator, accelerating the eastward current (Figure \ref{fign_formation_rossby}(b)), so that the effect of Rossby waves is also essential in the formation of the superrotation.

Besides the oceanic superrotation, the atmosphere is also in equatorial superrotation. However, atmospheric superrotation is not the direct cause of the oceanic superrotation especially if the equatorial zonal-mean wind stress is westward in some experiments with different rotation rates. This is due to the fact that the atmospheric superrotation concentrates in the free troposphere and rarely extends to the surface (see Figure 1 in \cite{del1996simulations}, Figure 2 in \cite{joshi1997simulations}, Figure 5 in \cite{read2018superrotation}, Figure 10 in \cite{yang2019ocean}, and Figure S6 in \cite{yang2020transition}). Near the surface, the direction and strength of the surface winds are mostly determined by surface temperature gradients (more exactly, the pressure gradient force) and frictions (\cite{lindzen1987role}). We have examined the zonal momentum budget for the surface atmosphere in experiments of varying rotation periods (8, 10, 30, 60, 80 and 100 days, which will be discussed in section \ref{sec:rotation}). In the rapid rotating experiments (rotation period of 8, 10 and 30 days), the dominating balance at low latitudes is a two-way balance between the pressure gradient force and friction, and the dominating balance at high latitudes is a three-way balance among the pressure gradient force, friction and Coriolis force; in the slow rotating experiments (rotation period of 60, 80 and 100 days), the dominating balance at the surface is between the pressure gradient force and friction in the whole planet (figure not shown). This is different from the momentum budget in the tropical Earth where the Coriolis force is also important (e.g., \cite{schneider2008eddy, faulk2017effects}), due to the slow rotation rates of tidally locked planets examined here.


\section{Maintenance mechanism} \label{sec:maintenance}

In equilibrium state, for quantifying the contributions of different processes to maintain the oceanic superrotation, we use the zonal- and temporal-mean zonal momentum equation:

\begin{eqnarray}\label{eq1}
\frac{\partial [\overline{u}]}{\partial t} &=& \frac{1}{\rho} \frac{\partial [\overline{\tau_x}]}{\partial z} + f[\overline{v}] + [\overline{F_x}] \nonumber \\
&-& \frac{[\overline{v}]}{a \cos{\phi}} \frac{\partial}{\partial \phi} ([\overline{u}\cos{\phi}]) - [\overline{w}] \frac{\partial [\overline{u}]}{\partial z} \nonumber \\
&-& \frac{1}{a \cos^2{\phi}} \frac{\partial}{\partial \phi} ([\overline{u}^* \overline{v}^* \cos^2{\phi}]) - \frac{\partial}{\partial z} [\overline{u}^* \overline{w}^*] \nonumber \\
&-& \frac{1}{a \cos^2{\phi}} \frac{\partial}{\partial \phi} ([\overline{u'v'} \cos^2{\phi}]) - \frac{\partial}{\partial z} [\overline{u'w'}],
\end{eqnarray}

\noindent where the brackets (asterisks) and overbars (primes) indicate zonal and temporal averages (deviations), respectively. The three terms in the first line are the wind stress, Coriolis force, and friction, respectively. The second to the forth lines are horizontal and vertical mean circulation terms, stationary eddies, and transient eddies, respectively; they are derived from nonlinear advection terms ($-u \partial u / \partial x -v \partial u / \partial y -w \partial u / \partial z$) by combining the incompressible continuity equation and taking zonal and temporal deviations (e.g., $uv = [\overline{u}][\overline{v}] + [\overline{u}^* \overline{v}^*] + [\overline{u'v'}]$). The advection terms do not influence the global momentum budget since their global integration is zero, but they represent momentum transport from one region to another region within the ocean. The left-hand-side term is close to zero in equilibrium state. This equation is the same as that used for diagnosing atmospheric supperrotation (\cite{kraucunas2005equatorial,lutsko2018response}), but with one additional term--the wind stress. 


\begin{figure}[ht]
\centering
\includegraphics[width=0.72\linewidth]{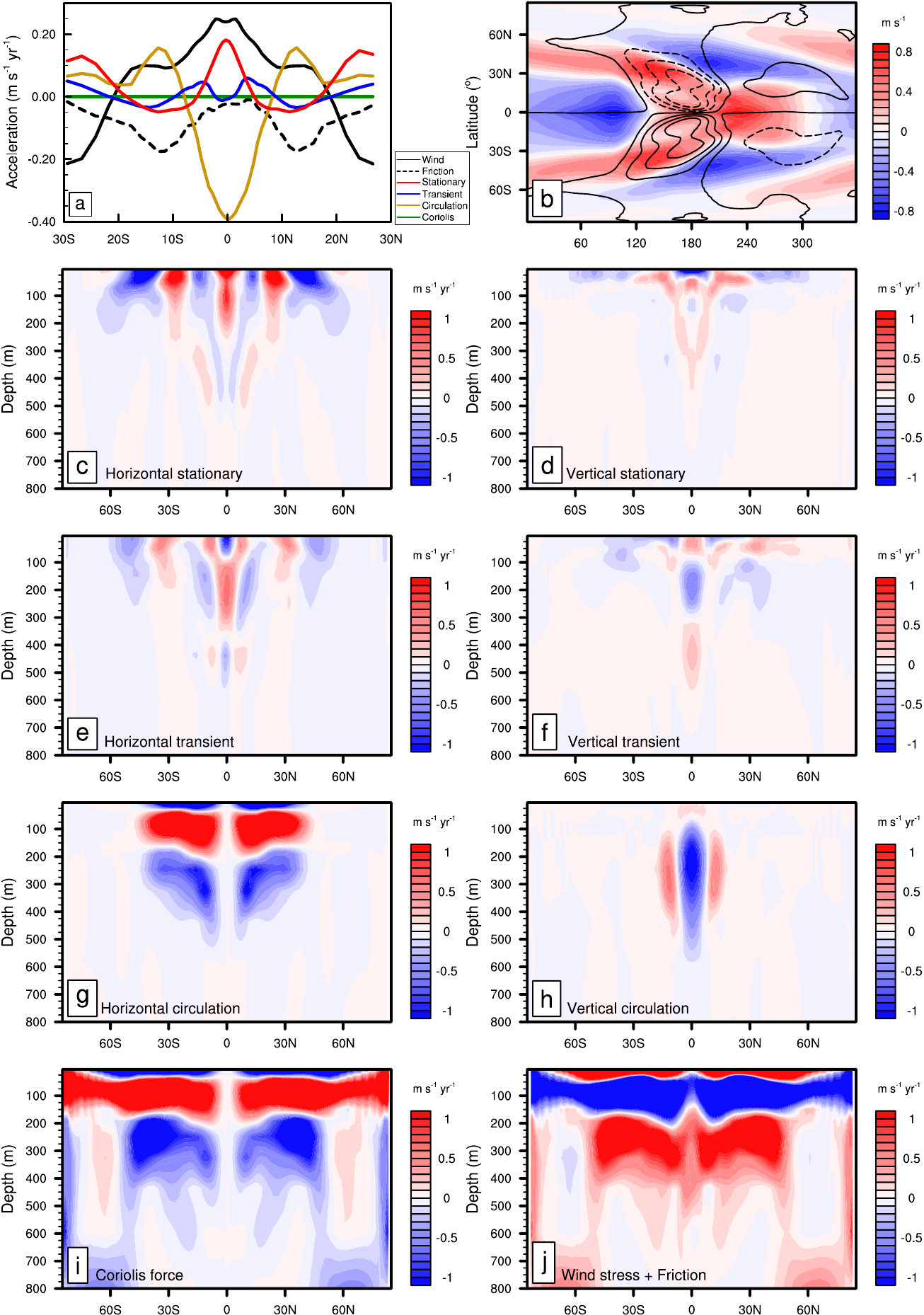}
\caption{Mechanism for sustaining the oceanic superrotation in the control experiment. (a): vertically-averaged zonal-mean zonal accelerations due to wind stress (black, solid), friction (black, dashed), stationary eddies (red), transient eddies (blue), mean circulation (yellow), and the Coriolis force (green) in the region between 30$^{\circ}$S and 30$^{\circ}$N. The ``Friction'' term is calculated by assuming the total acceleration to be zero because it is not outputted from the model. (b): stationary eddy velocities at the ocean surface as a function of longitude and latitude ($\overline{u}^*$ is color shaded, and $\overline{v}^*$ is line-contoured with an interval of 0.2~m\,s$^{-1}$ and with dashed lines representing negative values). (c) to (j): accelerations due to zonal-mean horizontal stationary eddies, vertical stationary eddies, horizontal transient eddies, vertical transient eddies, horizontal mean circulation, vertical mean circulation, Coriolis force, and wind stress and friction as a function of latitude and depth, respectively. The ``Wind stress + Friction'' term is calculated by assuming the total acceleration to be zero because friction is not outputted from the model. Note that the transient eddies with a time scale shorter than one month are also included in the residual terms in (j), because the shortest output time interval is one month, and the terms of $uv$ and $uw$ are not outputted from the model.}
\label{fig4_budget}
\end{figure}

The vertically-averaged zonal momentum budget in the equatorial region show that superrotation is maintained by stationary eddy momentum convergence combined with eastward zonal-mean wind stress (Figure~\ref{fig4_budget}(a)). When taking vertical average, the vertical components of the eddy terms vanish. We further treat the horizontal and vertical mean circulation terms as a whole. The friction term is calculated by assuming the total acceleration to be zero because it is not outputted from the model. The two accelerating terms are the wind stress and stationary eddies. The direction of the zonal-mean zonal wind stress is eastward, which gives a positive momentum input to the ocean. The stationary eddies are accelerating the equatorial current, which are caused by the coupled Rossby-Kelvin waves and will be discussed later in this section. These two terms together maintain oceanic superrotation and are balanced mainly by the mean circulation terms. The effects of transient eddies and friction are very small at the equator. The Coriolis force is zero everywhere because the vertically-averaged zonal-mean meridional velocity $[\overline{v}]$ is zero everywhere, due to the constraint of mass conservation.

For the vertical structure of the zonal momentum budget, at the equatorial region, the horizontal stationary eddy term accelerates the current between the surface and around 200~m, through transporting momentum from higher latitudes to the equator (Figure~\ref{fig4_budget}(c)). The vertical stationary eddy term transports momentum vertically from the upper 30~m ocean to the region between 30 and 100~m (Figure~\ref{fig4_budget}(d)). Unlike stationary eddies which is mainly constrained in the upper 200 m, the transient eddies are strong even at 450 m (Figure~\ref{fig4_budget}(e) \& (f)). However, the horizontal and vertical transient eddies cancel with each other so that the net effect of transient eddies is close to zero at the equator. The horizontal mean circulation term is close to zero in the equatorial region (Figure~\ref{fig4_budget}(g)) because the meridional velocity on the equator vanishes by symmetry, but it is strong at higher latitudes. The vertical mean circulation term acts to weakly accelerates the current between the surface and around 80~m but strongly decelerates the current between 80~m and seafloor, with the maximum at around 250~m (Figure~\ref{fig4_budget}(h)). This is because the strength of the zonal current decreases with depth (Figure~\ref{fig2_current}(b)) and the zonal-mean vertical velocity is upwelling in this experiment (Figure~\ref{fig2_current}(d)). This upwelling transports relatively lower-momentum seawater upward and thereby decelerates the current. The Coriolis force is small in the equatorial region but is one of the dominating forces at high latitudes (Figure~\ref{fig4_budget}(i)), and is mainly balanced by wind stress and friction there (Figure~\ref{fig4_budget}(j)). The structure of the Coriolis force, $f[\overline{v}]$, is similar to the horizontal mean circulation term, $-[\overline{v}] \partial [\overline{u}]/\partial y$ (compare Figure~\ref{fig4_budget}(g) \& (i)). This is because the relative vorticity, $-\partial [\overline{u}]/ \partial y$, is the same sign as the planetary vorticity $f$ (Figure~\ref{fig2_current}(b)). As a result, the patterns of these two terms are similar to the zonal-mean meridional velocity $[\overline{v}]$ in the northern hemisphere and reverses sign in the southern hemisphere. The patterns of the zonal-mean meridional velocity is similar to that found in \cite{hu2014role}. The converging current within around the surface 30~m is driven by converging wind stress towards equator, and the poleward currents between around 30~m to 200~m may be explained by a poleward Sverdrup transport. The converging ocean current in the deeper ocean is a compensation current as a result of mass conservation. The patterns of the ``Wind stress + Friction'' term are more complex. At the surface of the low-latitude ocean, it is positive due to positive zonal-mean wind stress. In other regions, the friction acts to close the budget. The mechanism of the three-layer spatial pattern is still unclear and requires future studies.

The stationary eddies are from the coupled Rossby and Kelvin waves in the ocean. These two wave modes exhibit eddy velocities ($\overline{u}^*$ and $\overline{v}^*$) tilting northwest-to-southeast on the northern hemisphere and southwest-to-northeast on the southern hemisphere (Figure~\ref{fig4_budget}(b)). This chevron-shape spatial pattern is similar to the Matsuno-Gill mode obtained in shallow water models (\cite{matsuno1966quasi,gill1980some}), but its horizontal scale is larger due to the slow rotation. The correlation between $\overline{u}^*$ and $\overline{v}^*$ leads to a southward (northward) momentum flux on the northern (southern) hemisphere, causing a momentum convergence to the equator and subsequently accelerating the current. This is similar to the process for maintaining the atmospheric superrotation on tidally locked hot Jupiters and terrestrial planets (\cite{showman2011equatorial,showman2013atmospheric}).

In a short summary, in the control experiment, the equatorial momentum is imported horizontally by stationary eddies from higher latitudes to the equator and meanwhile produced from surface winds; this momentum is vertically redistributed by eddies as well as internal stresses from the surface ocean to the interior of the ocean; the mean circulation, interior viscosity, and bottom friction act to consume the momentum. These processes act to maintain a steady state of the equatorial superrotation.


\section{The effect of planetary rotation rate} \label{sec:rotation}


\begin{figure}[t]
\centering
\includegraphics[width=0.95\linewidth]{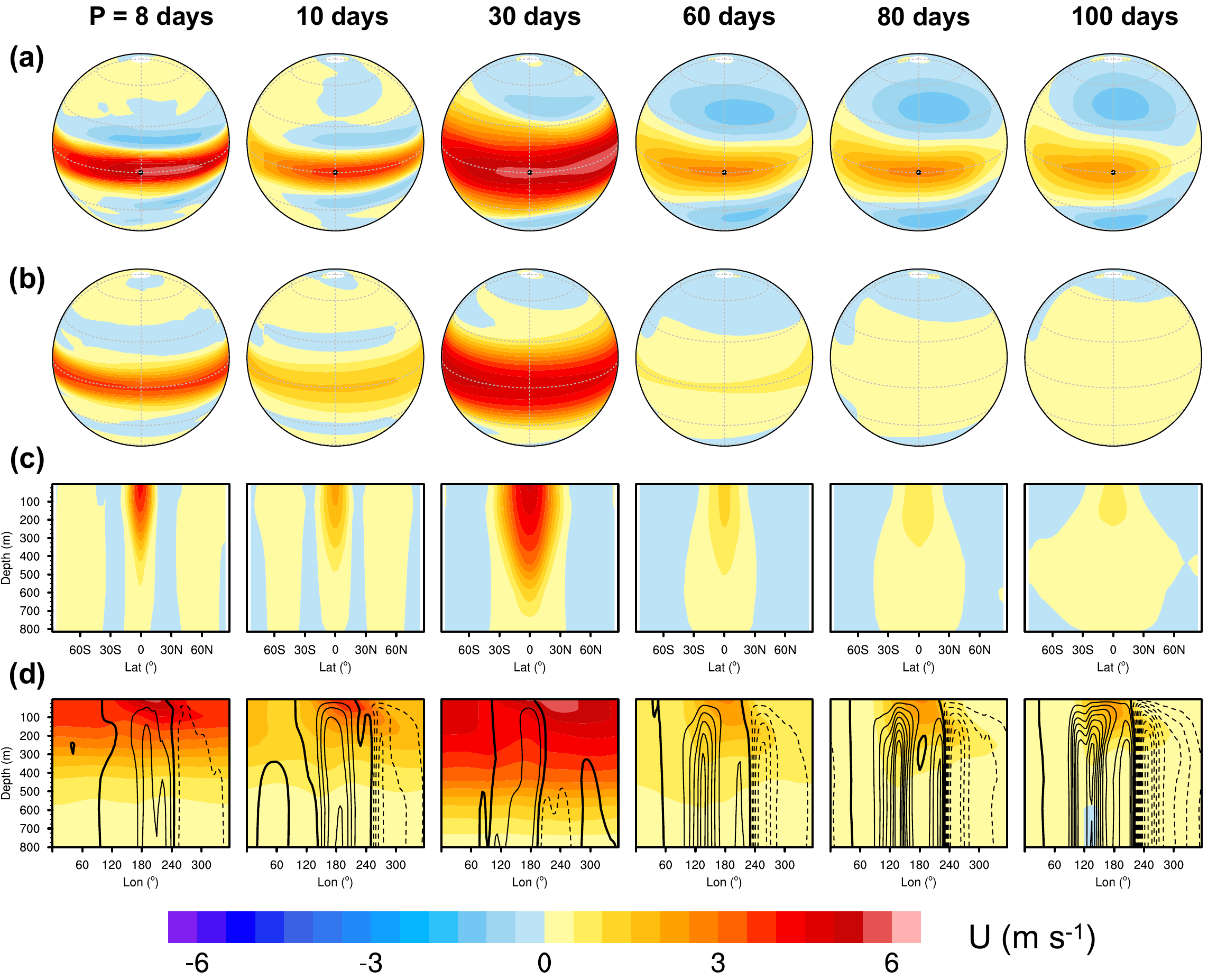}
\caption{Zonal velocity of the ocean under different planetary rotation periods from 8 to 100 Earth days. (a) and (b): surface zonal velocity on the day and night sides, respectively; (c): zonal-mean zonal velocity as a function of latitude and depth; and (d): zonal velocity along the equator (color shading) and the baroclinic component of pressure gradient force due to the spatial distribution of seawater density related to temperature and salinity differences (contour lines with an interval of 2 m s$^{-1}$ yr$^{-1}$; dashed lines are negative). The black dots in (a) indicate the substellar point. The stellar flux is 1400 W\,m$^{-2}$ and CO$_2$ concentration is 300\,ppmv in all these experiments.}
\label{fig5_rotation}
\end{figure}

\begin{figure}[b]
\centering
\includegraphics[width=0.85\linewidth]{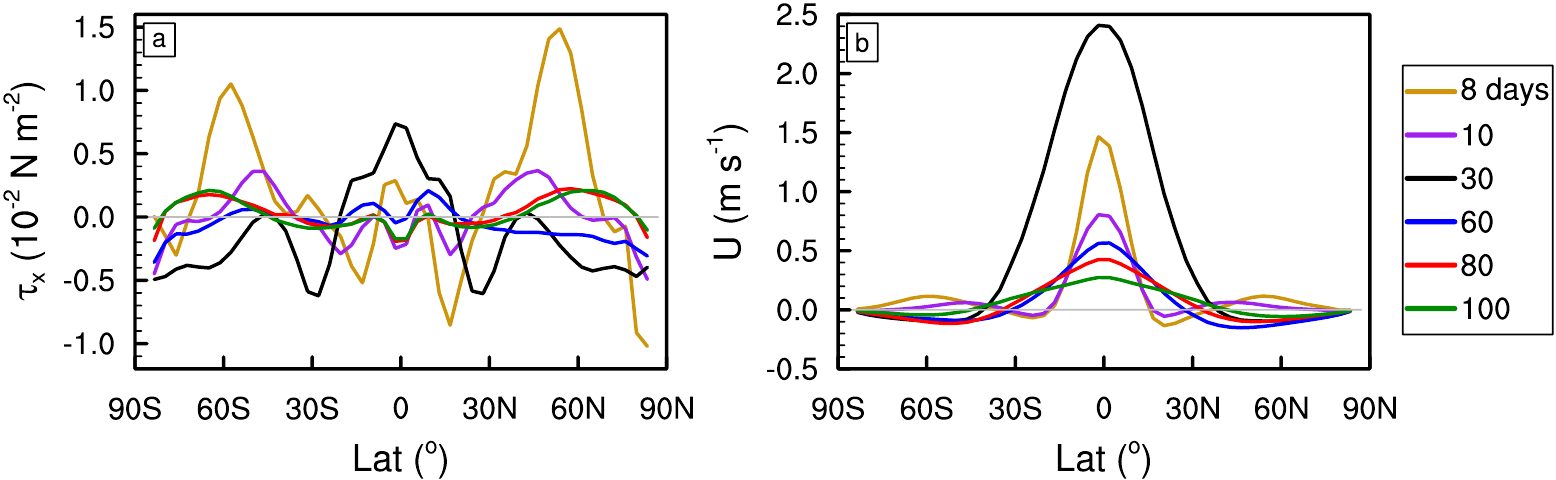}
\caption{Zonal-mean zonal surface wind stress (a) and vertically-averaged zonal-mean zonal ocean velocity (b) in the experiments of different planetary rotation periods. The equatorial zonal-mean zonal surface wind stress changes sign among these experiments, but oceanic superrotation exists in all the experiments.}
\label{figS_rotation_zonal}
\end{figure}

Planetary rotation rate influences the strength and width of the oceanic superrotation, as shown in Figure \ref{fig5_rotation} and Figure \ref{figS_rotation_zonal}(b). For relatively fast rotation rates (rotation period $=$ 8 or 10 days), the superrotation is strong but narrow, and there are several eastward and westward currents on each hemisphere, similar to that in the atmospheres of Jupiter and Saturn. The 30-day case also belongs to the rapid rotation planets, but it is special among the rapid rotation planets with the strongest superrotation, which may be because the 30-day case is ice-free. The strength of the superrotation is also dependent on different parameters like gravity, salinity, ocean-ice interaction and the parameterization scheme of oceanic mesoscale eddies (Figure \ref{figS_sensitivity}). How the strength of the oceanic superrotation is determined in detail, however, remain unclear due to the complexity brought by the coupling effects among the ice, ocean and atmosphere and the non-linear behavior of the system. For relatively slow rotation rates (rotation period $=$ 60, 80, and 100 days), the superrotation is robust on the day side but very weak on the night side. This is due to the interactions among atmospheric circulation, cloud, surface temperature, sea ice, and ocean.


\begin{figure}[b]
\centering
\includegraphics[width=1.0\linewidth]{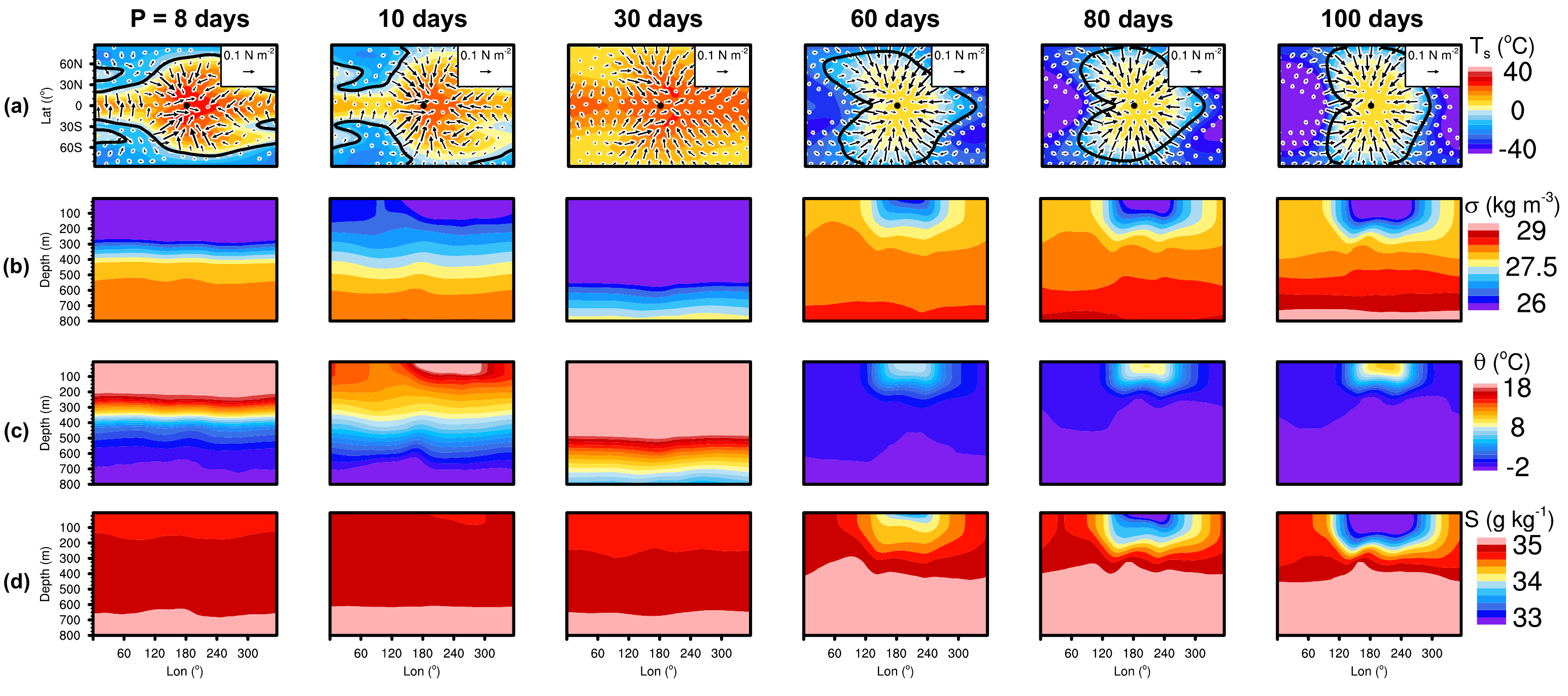}
\caption{Effects of planetary rotation rate on the climate, in the experiments of different rotation periods from 8 to 100 Earth days. (a) shows surface wind stress (vector) and surface air temperature (color shading). The black contour lines indicate the edge of sea ice defined by 50\% of sea ice concentration. The black dots indicate the substellar point. (b)-(d) show the potential density, potential temperature, and salinity along the equator, respectively. The planetary albedo is 0.36, 0.37, 0.30, 0.45, 0.46, and 0.48, and the surface albedo is 0.072, 0.069, 0.048, 0.11, 0.12, and 0.12 in the six experiments, respectively. The differences in the planetary albedo are most contributed from clouds; note that sea ice on the permanent night side has no albedo effect. In the three experiments of 60, 80, and 100 days, the night side is cooler for both surface and the underlying ocean, the day-night thermal contrast is greater, the spatial pattern is more symmetrical around the substellar point, and more sea ice forms on the night side. Sea ice formation on the night side increases the local seawater salinity through salt release, the ice is transported to the day side by surface winds, and sea ice melting on the day side decreases the local seawater salinity through freshwater input (\cite{yang2020transition}).}
\label{figS_slow_rotation}
\end{figure}


\begin{figure}[t]
\centering
\includegraphics[width=1.05\linewidth]{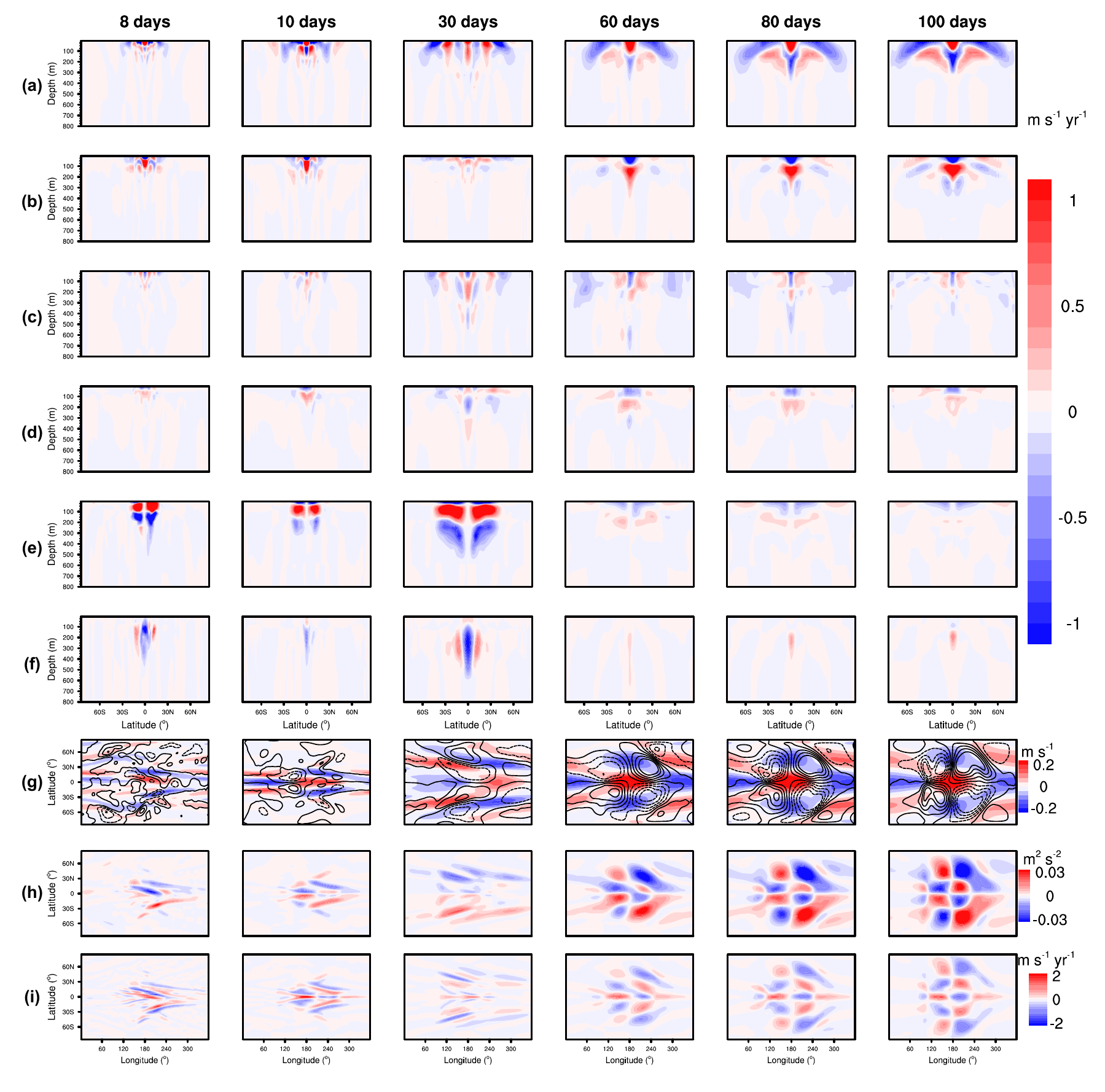}
\caption{Effects of eddies and mean circulation in the rotation rate experiments. In (a)-(f), they are accelerations due to zonal-mean horizontal stationary eddies (a), vertical stationary eddies (b), horizontal transient eddies (c), vertical transient eddies (d), horizontal mean circulation (e), and vertical mean circulation (f). For (g)-(i), they are vertically-averaged stationary eddy zonal velocity ($\overline{u}^*$, color shading) and meridional velocity ($\overline{v}^*$, contour lines with an interval of 0.02 m s$^{-1}$ and with dashed lines representing negative values, g), the vertically-averaged multiplication ($\overline{u}^*\overline{v}^*$, h), and the net acceleration ($-\partial \overline{u}^*\overline{v}^*/\partial y$, i), respectively. From left to right, the rotation periods are 8, 10, 30, 60, 80, and 100 Earth days, respectively. In all the experiments, the stationary eddies have a net effect of accelerating the equatorial current. The case of 30 days shown in panels (a)-(f) is the same as that shown in Figure \ref{fig4_budget}(c)-(h) for comparison. Note that in the equilibrium state, the magnitude of acceleration and deceleration (i) is about one order smaller than that during the formation phase (see Figure \ref{fig3_generation}(f)).}
\label{figS_eddies1}
\end{figure}

For slowly rotating planets, atmospheric upwelling over the substellar region is stronger, and the cloud albedo is higher (Figure \ref{figS_timeseries}(c), see also \cite{merlis2010atmospheric,yang2013stabilizing,carone2015connecting,Kopparapu2017the,Noda2017the,haqq-misra2018demarcating}), so that the surface temperature is lower on both day and night sides; the atmospheric superrotation is also weaker, so that the day-to-night surface temperature contrast is larger, and more sea ice forms on the night side (Figure \ref{figS_slow_rotation}). Cooler seawater and higher salinity from brine rejection during the sea ice formation make the seawater on the night side denser than that on the day side, inducing a negative (positive) baroclinic component of the pressure gradient force around the east (west) terminator (contour lines in Figure \ref{fig5_rotation}(d)). Furthermore, the stronger surface winds at the east terminator (vectors in Figure \ref{figS_slow_rotation}(a)), combined with the strong density contrast, act to prevent the propagation of Kelvin waves to the night side. These two processes cause the superrotation to be trapped on the day side in the three slow rotation experiments.

The zonal-mean surface wind stress at the equator is not eastward in all the experiments. In the experiments of 10, 80, and 100 days, it is close to zero or westward rather than eastward (Figure~\ref{figS_rotation_zonal}(a)), in contrast to that in the control experiment. However, oceanic superrotation does exist and horizontal eddy momentum flux is equatorward in all the experiments (Figure~\ref{figS_eddies1}(a)). These suggest that an equatorial zonal-mean eastward wind stress is not necessary for maintaining the superrotation while an equatorward eddy momentum flux is. Note that this does not indicate wind stresses are unimportant. The Rossby and Kelvin waves are generated from the undulations of the sea surface height by the surface winds as well as by the thermal forcing from the stellar radiation.

Planetary rotation rate also affects the patterns of eddies in the ocean. The effects of stationary eddies are the same in all simulations: the horizontal stationary eddies transport momentum equatorward from higher latitudes, and the vertical stationary eddies transport momentum downwards (Figure \ref{figS_eddies1}(a) \& (b)). As a result, the mechanisms of the maintenance of oceanic superrotation are robust when varying rotation rates. As rotation rate decreases, the spatial scale of the horizontal stationary eddies increases due to the increased Rossby deformation radius (Figure \ref{figS_eddies1}(g)-(i)). Among the rapid rotating planets (8, 10 and 30 days), the vertical stationary eddies are weakest in the 30-day case. Among the slow rotating planets (60, 80 and 100 days), as rotation rate decreases, the horizontal scale of vertical stationary eddies gets larger due to the increased Rossby deformation radius, while the vertical extension gets shallower (Figure \ref{figS_eddies1}(b)). The spatial pattern and strength of transient eddies also change with rotation rates, but they are relatively weak compared with stationary eddies (Figure \ref{figS_eddies1}(c) \& (d)).

The mean circulation is also strongly affected by the rotation rates. Generally, oceanic circulation is strong in the rapid rotating planets but is weak in the slow rotating planets. As a result, the mean circulation terms are important in the rapid rotating planets but are small in the slow rotating planets. The spatial scale of the horizontal mean circulation also increases as rotation rate decreases (Figure \ref{figS_eddies1}(e)) due to the increased Rossby deformation radius. The horizontal mean circulation term is close to zero at the equator because of vanishing zonal-mean meridional velocity $[\overline{v}]$ due to symmetry. At subtropical and extratropical regions, the mean circulation terms are caused by meridional flows that advect higher/lower momentum water poleward/equatorward, which is similar to that in the 30-day case. There is almost no ``compensation'' converging current in the deep ocean in the slow rotating planets, which may result from weak meridional transport in the upper ocean. At the equator, the vertical mean circulation acts to dissipate momentum in the rapid-rotating planets, but is weakly accelerating the ocean in the slow-rotating planets (Figure \ref{figS_eddies1}(f)). This is because the equatorial vertical motion in the rapid rotating planets is dominated by upwelling which transports lower momentum water upwards, but is dominated by downwelling in the slow rotating planets which transports higher momentum water downwards.


\begin{figure}[b]
\centering
\includegraphics[width=0.95\linewidth]{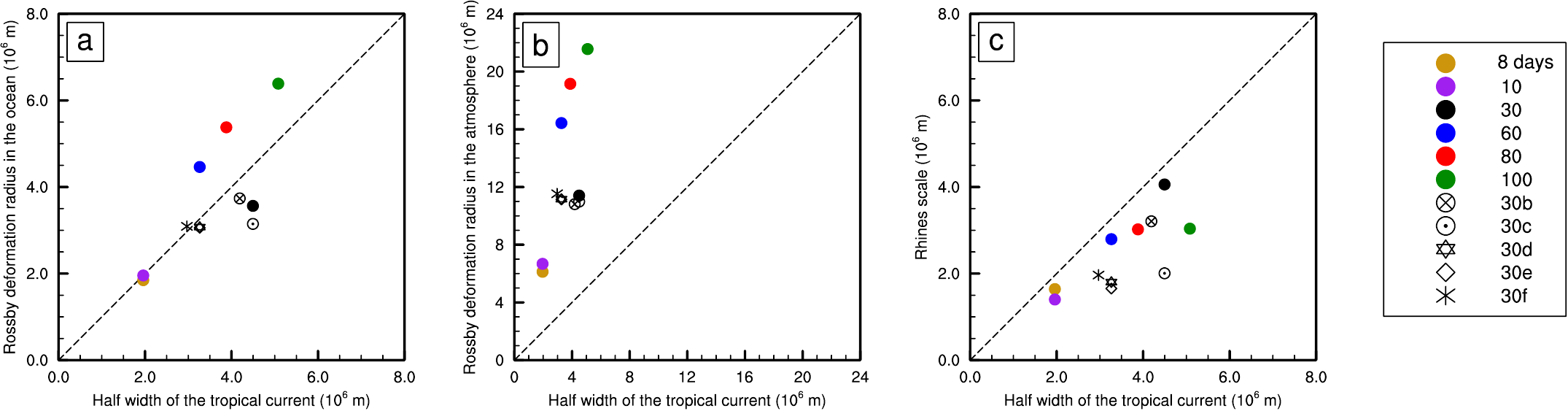}
\caption{Scaling for the width of the tropical current. The $y$-axis indicates Rossby deformation radius in the ocean in (a), Rossby deformation radius in the atmosphere in (b), and the Rhines scale in the ocean in (c), respectively. The horizontal axis is the half width of the tropical current in all three panels, defined as the latitude where the vertically-averaged zonal-mean ocean zonal velocity is zero. The dashed lines are the one-to-one lines. Experiments include the rotation rate cases (8, 10, 30, 60, 80, and 100 days) and the sensitivity experiments (30b, 30c, 30d, 30e, and 30f, corresponding the cases shown in the Figure \ref{figS_sensitivity}(b)-(f), respectively). The rotation period is 30 days in all the sensitivity experiments.}
\label{fig6_scalings}
\end{figure}

The width of the superrotation increases with longer rotation period, which indicates larger Rossby deformation radius. This trend is the same as the trend of the horizontal scale of the atmospheric circulation when varying rotation rates (\cite{showman2013atmospheric, kaspi2015atmospheric, guendelman2019atmospheric}). In our simulations, the width of the equatorial current well scales with the Rossby deformation radius in the ocean (Figure \ref{fig6_scalings}(a)). But, the Rossby deformation radius in the atmosphere is around 3-4 times of the current's width and the Rhines scale in the ocean is smaller than the current's width in all the experiments (Figure \ref{fig6_scalings}(b)-(c)). The Rossby deformation radius ($L_R$) is the length scale at which rotational effects become as important as the effects of gravity waves or buoyancy in the evolution of the flow in a disturbance. In the tropics, $L_R$ is equal to $(NH/\beta)^{1/2}$, where $N$ is the Brunt-V\"ais\"al\"a frequency, $H$ is the scale height for the atmosphere or the depth of the current for the ocean, and $\beta$ is the northward gradient of the Coriolis parameter ($f$). The Rhines scale is a transition length between the regime dominated by small-scale 2D turbulence and the regime dominated by unforced Rossby waves (\cite{rhines1975waves, rhines1979geostrophic}). It is equal to $\pi (2U_{rms}/\beta)^{1/2}$, where $U_{rms}$ is the root mean square velocity. These results indicate that in maintaining the oceanic superrotation, forced Rossby waves in the ocean is the key, rather than Rossby waves in the atmosphere or free Rossby waves in the ocean. 

\begin{figure}[b]
\centering
\includegraphics[width=0.8\linewidth]{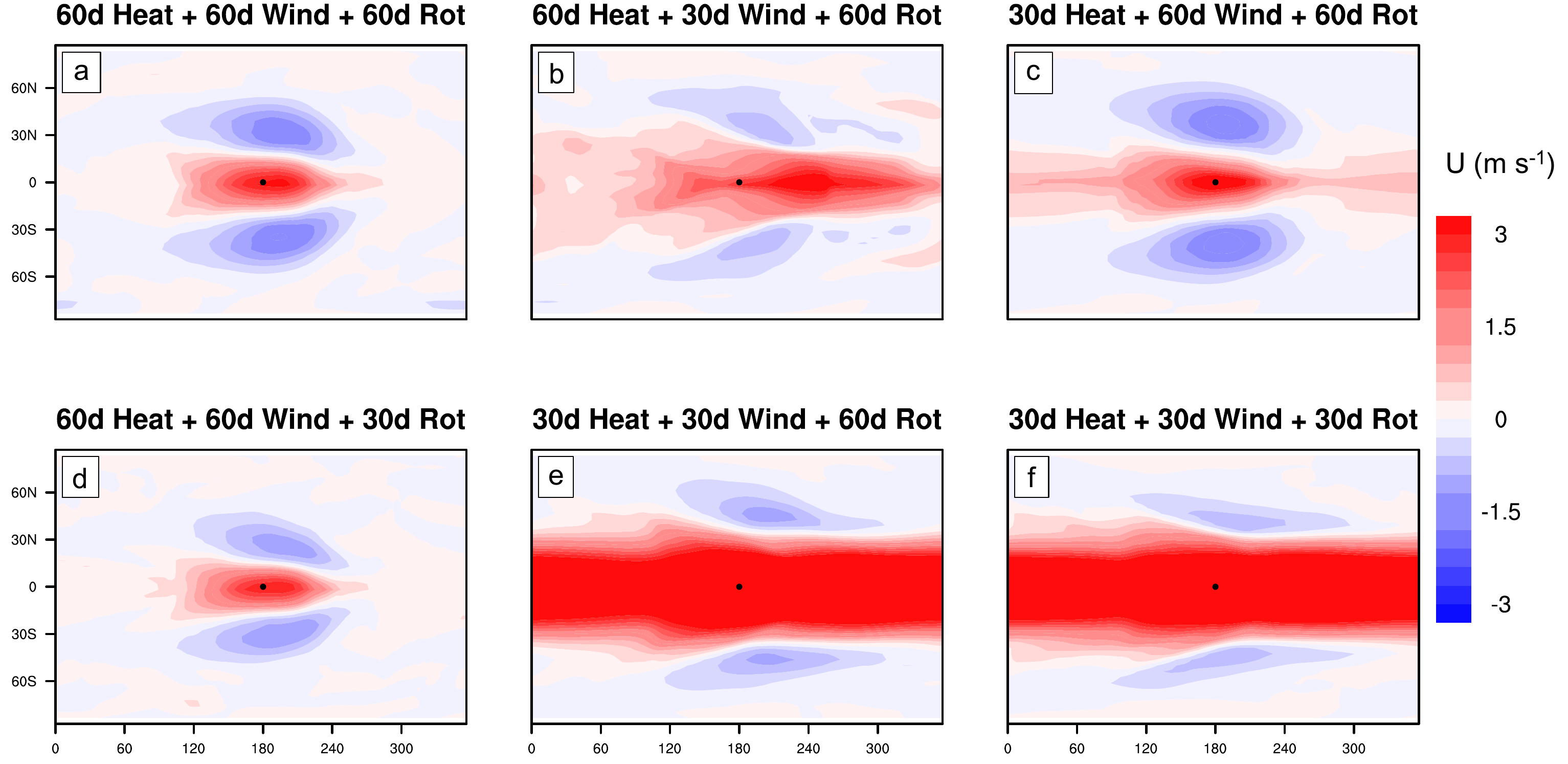}
\caption{Effects of surface wind stresses, surface heat fluxes, and rotation rate on the oceanic superrotation. (a): oceanic surface zonal velocity in the ocean-only experiment with specified surface wind stresses and heat fluxes from the coupled atmosphere-ocean experiment of 60 days and with a rotation period of 60 days; (b): same as (a) but with the wind stresses replaced by those from the coupled experiment of 30 days; (c): same as (a) but with the heat fluxes replaced by those from the coupled experiment of 30 days; (d): same as (a) but with the rotation period set to 30 days; (e): same as (a) but with both heat fluxes and wind stresses replaced by those from the coupled experiment of 30 days; (f): same as (e) but with the rotation period further set to 30 days. Black dots in (a)-(f) indicate the substellar point. These experiments suggest that both wind stresses and heat fluxes are important in influencing the strength of the oceanic superrotation while the Coriolis force in the ocean is important in affecting its spatial scale.}
\label{figS_pure-ocean1}
\end{figure}

In order to further understand the separate effects of the Coriolis force, surface wind, and surface heat flux on the oceanic superrotation, we performed ocean-only experiments using the ocean component of the model (see Section \ref{sec:Methods}). The ocean-only simulations forced with surface winds and surface fluxes can well reproduce the results of the coupled model in the 60-day and 30-day simulations (see Figure \ref{figS_pure-ocean1}(a) versus the 4$^{th}$ column in Figure \ref{fig5_rotation}, and Figure \ref{figS_pure-ocean1}(f) versus Figure \ref{figS_sensitivity}(a)). We further carry out simulations by changing the forcing of the surface winds, heat fluxes and rotation periods one by one from the 60-day case to the 30-day case to study the evolution of oceanic circulation and the intensification of the oceanic superrotation.


When only the surface wind is replaced from the 60-day case to the 30-day case, the superrotation gets stronger at the equator (Figure \ref{figS_pure-ocean1}(b)). This is because the zonal-mean zonal wind stress is stronger in the latter case (Figure \ref{figS_rotation_zonal}(a)). However, the strength of the superrotation is still much weaker than that in the 30-day case, especially on the west of the substellar point.

When only the surface heat flux is replaced from the 60-day case to the 30-day case, the equatorial jet is intensified on the night side (Figure \ref{figS_pure-ocean1}(c)). This is because in the 30-day case, the day-night surface temperature contrast is smaller (Figure \ref{figS_slow_rotation}(a)). Smaller day-night thermal forcing contrast results in smaller density contrast in the ocean, so that the Kelvin wave is easier to propagate to the night side hence the equatorial eastward jet gets stronger on the night side.

When only the rotation period is replaced from the 60-day case to the 30-day case, the strength of the ocean current does not change significantly but its spatial scale becomes relatively smaller (Figure \ref{figS_slow_rotation}(d)). This is because when the rotation rate gets faster, the Rossby deformation radius decreases, so that the spatial scale of ocean currents gets smaller.

When both the surface winds and heat fluxes are replaced with those from the coupled experiment of 30 days, the ocean current is very similar to that in the 30-day simulations, except for small differences in the spatial scales due to different rotation rates. These results imply that both wind stress and thermal forcing influence the strength of oceanic superrotation, and the combined effect is far more significant than each separate effect (comparing Figure \ref{figS_pure-ocean1}(a), (b), (c) and (e)). The Coriolis force does not affect the strength of oceanic superrotation significantly, but it acts to constrain the spatial scale of the oceanic superrotation through determining the Rossby deformation radius in the ocean.


\begin{figure}[t]
\centering
\includegraphics[width=0.85\linewidth]{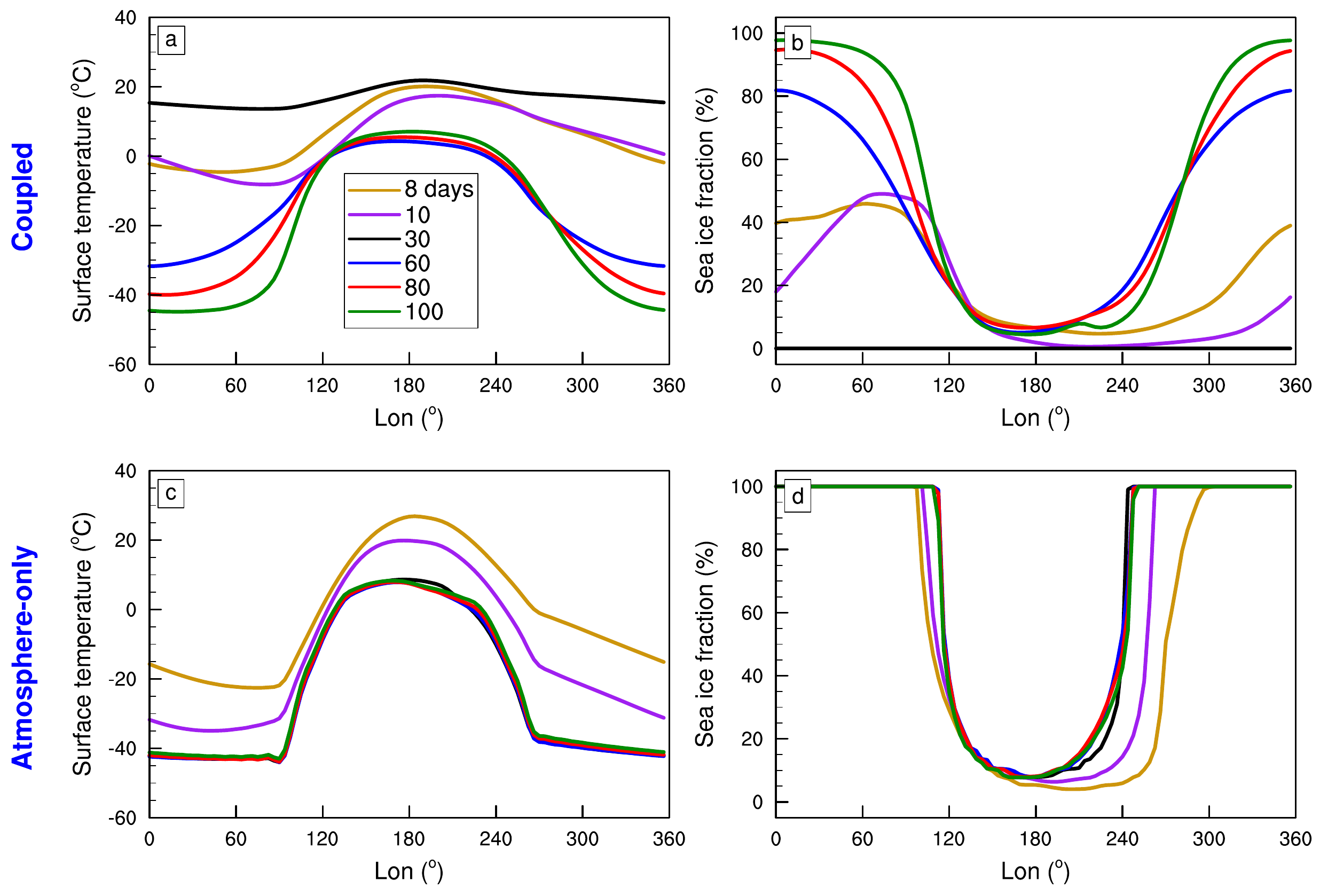}
\caption{Climatic effects of the oceanic superrotation in the experiments of different rotation periods. (a) \& (b): meridional-mean (south-north-mean) surface air temperature and sea ice fraction, respectively, in the coupled atmosphere-ocean experiments. (c) \& (d): same as (a) \& (b) but for the atmosphere-only experiments coupled to a slab ocean with no oceanic heat transport. Oceanic dynamics are effective in warming the night side in the experiments of 8, 10, and 30 days.}
\label{fig7_climatic}
\end{figure}

Besides of the rotation rate and the parameters shown in Figure~\ref{figS_sensitivity}, other factors can also influence the strength of the oceanic superrotion, such as greenhouse gas concentration, ocean depth, and stellar flux investigated in \cite{yang2014water,yang2019ocean}. As shown in Figure~1 of \cite{yang2014water} and Figure~12 of \cite{yang2019ocean}, the superrotation weakens and concentrates on the day side when the ocean depth reduces. This is due to the increased effect of ocean bottom drag. Moreover, \cite{yang2019ocean} showed that the superrotation becomes much weaker for planets close to the inner edge of the habitable zone. This is due to the small temperature differences between day and night side and subsequently very weak surface winds everywhere in hot climates.

The oceanic superrotation can influence planetary climate. In order to clearly know the climatic effects of oceanic dynamics, we performed the same experiments but with an atmosphere-only model (see section \ref{sec:Methods}). In the slowly rotating cases (60, 80, and 100 days), there is no large difference between these two types of experiments (coupled atmosphere-ocean and atmosphere-only), indicating that oceanic heat transport to the night side is not effective on the slowly rotating planets (Figure \ref{fig7_climatic}). In the fast rotating cases (8, 10 and 30 days), most of the night-side sea ice melts and the surface temperature is quite uniform along the longitude in the coupled experiments, whereas in the atmosphere-only experiments the whole planet is cooler and the entire night side is covered by sea ice. This is related to the strong circumequatorial oceanic superrotation, which transports heat from the substellar region to the night side, reducing the day-to-night thermal contrast, decreasing the cloud albedo, and warming the surface especially on the night side (\cite{hu2014role, yang2019ocean}). The ice albedo feedback is weak on tidally locked planets. This is because sea ice fraction is close to zero on the day side in all simulations (Figure~\ref{figS_slow_rotation}(a) \& Figure~\ref{fig7_climatic}(b)), and there is no stellar radiation on the night side. In these six simulations, the planetary albedo varies from 0.30 to 0.48, and the surface albedo varies from 0.05 to 0.12, so that sea ice has smaller contribution to the planetary albedo than clouds.


\section{Summary and Discussions} \label{sec:summary}

In this study, we uncover the formation and maintenance mechanisms of oceanic superrotation on tidally locked planets, and its sensitivity to different parameters especially the planetary rotation rate. The main results are as follows:

\begin{figure}[b]
\centering
\includegraphics[width=0.85\linewidth]{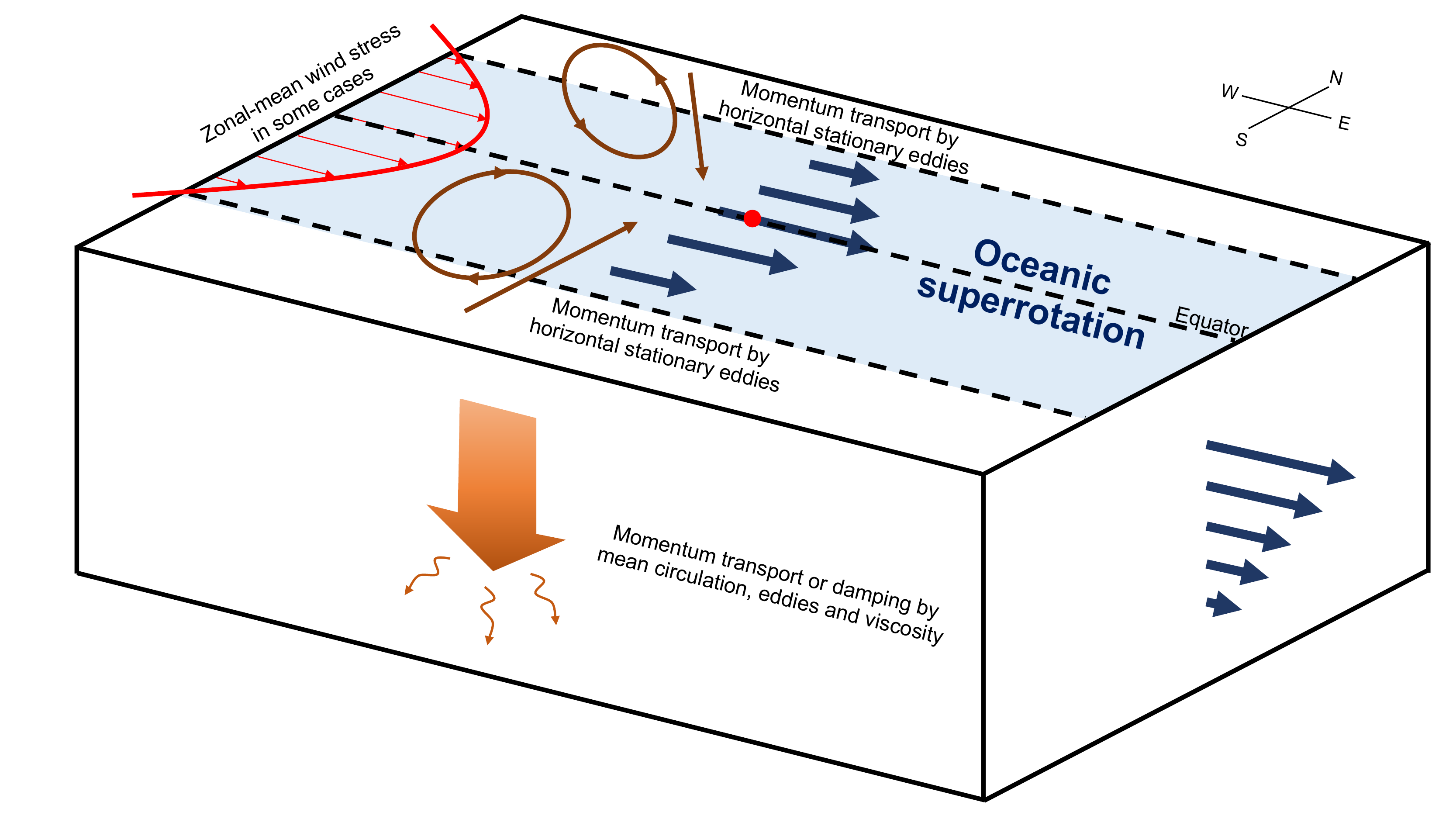}
\caption{Schematic diagram for the maintaining mechanism of the oceanic superrotation on tidally locked planets. The superrotation is maintained by equatorward momentum transports associated with stationary Rossby-Kelvin waves in the ocean, as well as in some cases by eastward momentum productions from zonal-mean eastward wind stresses. Red dot indicates the substellar point.}
\label{fig1_summary}
\end{figure}

\begin{enumerate}
    \item The generation of oceanic superrotation is caused by equatorial Kelvin waves that propagate eastward from the substellar point, as well as low-latitude Rossby waves that converges eastward momentum to the equator, and in some cases by the combined effect of an eastward zonal-mean zonal wind.
    \item The maintenance of oceanic superrotation is mainly driven by stationary Rossby-Kelvin waves as summarized in Figure \ref{fig1_summary}, which are excited from sea surface height disturbances that are associated with the uneven stellar energy distribution and the convergence of surface winds towards the substellar point. In some cases, the eastward zonal-mean wind stress at the equator also directly contributes to the oceanic superrotation.
    \item The width of the superrotation is constrained by the Rossby deformation radius in the ocean rather than that in the atmosphere, and the strength of the superrotation is influenced by the combined effects of wind stress, seawater salinity, and day-to-night thermal contrast, as well as the climate feedback associated with cloud, surface temperature, and ice formation.
    \item Under given stellar radiation, for planets of rapid rotation rates, the generally stronger superrotation significantly enhances the day-to-night oceanic heat transport, warms the night side and increases the planetary habitability, but for planets of very slow rotation rates, the climatic effect is smaller.
\end{enumerate}

There are many similarities and differences between the superrotation in the atmosphere and the ocean on tidally locked planets. The formation and maintenance mechanisms for superrotation are similar in both the atmosphere and the ocean: stationary eddy momentum convergence which are associated with the coupled Rossby-Kelvin waves near the equator, rooted in the strong day-night stellar radiation contrast. Besides, in both the atmosphere and the ocean, the width of the superrotation is mainly determined by the Rossby deformation radius near the equator, which is associated with the planetary rotation rate and the spherical geometry effect of the planet ($\beta$-effect). The Rossby deformation radius is larger in the atmosphere than in the ocean mainly because the scale height in the atmosphere is larger than the depth of the ocean. The strength of the superrotation is stronger in the atmosphere (O(10-100 m s$^{-1}$)) than that in the ocean (O(1 m s$^{-1}$)) due to smaller viscosity. Furthermore, superrotation in both the atmosphere and the ocean can transport heat from the day side to the night side and result in the eastward shift of the hottest spot. Although the strength of the equatorial jet is weaker in the ocean than in the atmosphere, the day-to-night heat transport in the ocean can be stronger than that in the atmosphere if there is no land blocking and the ocean current is significant (see Figure 3 in \cite{hu2014role}), due to large heat capacity of the ocean. Generally, day-to-night heat transport on tidally locked planets is dominated by the atmosphere near the inner edge of the habitable zone and is dominated by the ocean with smaller stellar fluxes (\cite{yang2019ocean}). Meanwhile, the atmospheric superrotation concentrates in the middle and upper layer of the troposphere and the oceanic superrotation is strongest near the surface.

The coupling between the atmospheric and oceanic superrotation mainly lies in two aspects: direct momentum connections and indirect thermodynamic connections. First, the surface wind field drives the oceanic circulation, and the surface drag from the ocean also influences the atmospheric circulation, representing the direct momentum exchange between the atmosphere and the ocean. Second, evaporation and precipitation connect the source and sink of the water vapor in the atmosphere and the salinity in the ocean. The latent heat associated with the phase change of water vapor is important in the atmospheric circulation, and the density variation associated with the salinity change is essential in the oceanic circulation, representing the indirect thermodynamic connection between the atmosphere and the ocean. Note that the phase change of the water vapor is also associated with cloud formation and the salinity change is also associated with ice formation and melting, and the further climatic effects such as albedo feedback increases the complexity of the system.

In this study, we show a clear picture of how oceanic superrotation is generated and maintained on tidally locked planets (Section \ref{sec:formation} \& \ref{sec:maintenance}). There are still some problems that require future studies. First, there exhibits non-monotonic dependence on rotation rates for oceanic circulation. We have discussed the differences between the rapid rotating planets and the slow rotating planets, and contributed the non-monotonic behaviour within the rapid rotation planets to the coupling and feedback associated with ice, but the detailed mechanisms remain unclear. Second, we have examined how rotation rate influences oceanic superrotation, and found that the general spatial scale is relevant to the Rossby deformation radius. How the strength and detailed spatial patterns are influenced by rotation rate and other planetary parameters remains unclear, though. Third, we have used ocean-only simulations to separate the effect of wind stresses, thermal fluxes and rotation rate on the oceanic superrotation. We have had some general discussions about the separate effects of these three components, but their effects in detail, however, still require further studies (see Section \ref{sec:rotation} for more detailed discussions).

Here we focus on planets in 1:1 tidally locked orbits. For planets in spin-orbit resonances (e.g., \cite{del2019habitable, ohno2019atmospheresI, ohno2019atmospheresII}) or Earth-like non-tidally locked orbits (e.g., \cite{kaspi2015atmospheric, guendelman2019atmospheric}), there can exist atmospheric superrotation, but whether and how oceanic superrotation can exist on these planets require future studies. Besides, in this study, we analyze the ocean dynamics on tidally locked planets which are ice-free or partially covered by sea ice. Future work can focus on the ocean dynamics and whether oceanic superrotation can exist on tidally locked planets that are in snowball state, on which the effect of wind stress is weak or negligible (e.g., \cite{ashkenazy2013dynamics, ashkenazy2014ocean, ashkenazy2016variability}). Moreover, tidally locked simulations here focus on relatively shallow oceans. If the depth of the ocean is tens of kilometers, deep oceanic circulation is required to be taken into consideration seriously. In addition, we set the atmosphere compositions to be Earth-like. Different atmospheric compositions can influence horizontal temperature gradients and subsequently the strength of surface winds, which should be investigated in the future. Further more, future work could also examine tidally locked oceans of different compositions, such as methane oceans on planets far from the host stars and magma oceans on planets close to the host stars. When the composition of the liquid ocean is changed, the physical properties, such as density, viscosity, diffusivity, and freezing point of the fluid can change, and a completely new scenario of ocean dynamics may occur.


\acknowledgments
We are grateful to the helpful discussions with Xueyang Zhang, Yonggang Liu, Malte Jansen, Xianyu Tan, Adam Showman, Shineng Hu, and Xiaozhou Yang. J.Y. was supported by the Chinese National Science Foundation under grants 42075046, 41675071, and 41861124002.

\bibliography{oceansuperrotation_v2}{}
\bibliographystyle{aasjournal}

\end{document}